\documentclass{aa}

\usepackage{graphicx}
\usepackage{txfonts}
\usepackage{longtable,lscape}
\begin{document}

    \title{VLT multi-epoch radial velocity survey toward \object{NGC 6253} 
     \thanks{Based on observations collected at the European Organization for Astronomical Research 
             in the Southern Hemisphere (ESO) in Paranal during
             program 381.C-0270(A), and in La Silla during program 073.C-0227(A), 
             and program 083.A-9001(C) thanks to Max Planck Institute for Extraterrestrial
             Physics reserved time.}}

    \subtitle{Analysis of three transiting planetary candidates}

  \author{M. Montalto
          \inst{2,1},
          S. Villanova
          \inst{3},
          J. Koppenhoefer
          \inst{1,2}
          }

   \authorrunning{M. Montalto et al.}

   \offprints{M. Montalto,\\
              \email{montalto@usm.uni-muenchen.de} }

   \institute{Universitaets-Sternwarte  der
              Ludwig-Maximilians-Universitaet, Scheinerstr.1, 81679 Muenchen, Germany.
              \and
              Max-Planck-Institute for Extraterrestrial Physics, Giessenbachstr.,
              Garching b Muenchen, 85741, Germany.
             \and
              Grupo de Astronomia, Departamento de Fisica, Casilla 160, Universidad de Conception, Chile
             }

   \date{}

\abstract
{
}
{
We measured the radial velocity of 139 stars in the region of \object{NGC 6253},
discussing cluster's membership and binarity in this sample, 
complementing our analysis with photometric, proper motion, and
radial velocity data available from previous studies of this cluster, 
and analyzing three planetary transiting candidates we found in the field
of \object{NGC~6253}.
}
{
Spectra were obtained with the 
UVES and GIRAFFE spectrographs at the VLT, during three epochs in August 2008.
}
{
The mean radial velocity of the cluster
is ($\overline{RV_{cl}}\pm\overline{\sigma_{cl}})=(-29.11\pm0.85)$ km/s. 
Using both radial velocities and proper motions we found
35 cluster's members, among which 12 are likely cluster's close binary
systems. One star may have a sub-stellar companion, requiring a more
intensive follow-up.
Our results are in good agreement with past radial velocity and photometric measurements. 
Furthermore, using our photometry, astrometry and spectroscopy we identified a new sub-giant 
branch eclipsing binary system, member of the cluster. The cluster's close binary 
frequency at (29$\pm$9)$\%$ (34$\%$$\pm$10$\%$ once including 
long period binaries), appears higher than the field binary frequency equal to (22$\pm$5)$\%$,
though these estimates are still consistent within the uncertainties.
Among the three transiting planetary candidates the brightest one ($V=15.26$) is worth to be more intensively investigated
with higher percision spectroscopy.
}
{
We discussed the possibility to detect sub-stellar companions (brown dwarfs and planets) with the radial
velocity technique (both with UVES/GIRAFFE and HARPS) around turn-off stars of old open clusters. 
We isolated 5 stars that are optimal targets to search for planetary mass companions
with HARPS. Our optimized strategy minimizes the observing time requested to isolate and follow-up best planetary
candidates in clusters with high precision spectrographs, an important aspect given the faintness of the target stars. 
} 


   \maketitle

%

\section{Introduction}
\label{s:introduction}


The advent of multi-object spectrographs feeding large aperture telescopes 
offers the possibility to obtain simultaneous, repeated, and accurate spectroscopic 
measurements of a large sample of stars. The application of 
this technique to the study of open clusters 
is of particular interest. While open clusters photometric
surveys are now routinely performed spanning typically 
several consecutive nights, there are relatively few multi-epoch radial velocity
surveys targeting these objects (Mermilliod et al. 2009; Hole et al. 2009). 
Besides allowing further culling 
of cluster's members, multiple radial velocity measurements 
allow a more complete census of binary and multiple systems.
Spectroscopic follow-up of eclipsing binary systems 
that belong to open clusters gives the opportunity to derive accurate stellar masses,
constraining models of stellar evolution (e. g. Southworth \& Clausen 2006). 
Moreover, the increasing stability and accuracy of modern 
spectrographs is driving toward detection of binary systems with low 
mass companions (such as brown dwarfs and planets) also in open clusters, 
where target stars are typically fainter than those observed in common 
radial velocity planet searches in the solar surrounding.

In this work, we focused our attention on the old and metal-rich open cluster
\object{NGC 6253} ($\alpha_{2000}=16^{h}\,59^{m}\,05^{s},
\delta_{2000}=-52^{\circ}\,42\arcmin\,30\arcsec, l=335\fdg5,b=-6\fdg3$). 
This cluster was selected as part
of our project aimed at searching for transiting hot-jupiter planets
in metal-rich open clusters (Montalto et al. 2007). In 2004, we performed
a photometric transit search toward \object{NGC 6253} using the WFI at the 2.2m
Telescope for ten consecutive nights (Montalto et al. 2009), 
identifying three transiting planet candidates. 

In the solar neighborhood, FGK metal-rich dwarf stars have higher probability 
to host jupiter-like planets (
Gonzalez 1998, Santos et al.~2001, 
 Fischer \& Valenti 2005). Accordingly to the 
most accredited explanation that the planet-metallicity 
correlation is of primordial origin, we expect
a higher planet discovery rate targeting preferentially
stars borned in metal-rich environments. 
While at solar metallicity the frequency of planets
(with period P $<$ 4 years, and radial velocity
semi-amplitude $K>30$ m/s) around FGK dwarf 
stars in the solar neighbourhood is about 3$\%$,
at the metallicity of \object{NGC~6253}, the 
expected frequency is around $18\%$, as can be deduced from
Fischer \& Valenti (2005). 

Since the large expected frequency of planets in metal rich clusters,
these objects are top targets for planet searches and offer an 
excellent natural laboratory to test ideas of planet formation
and evolution, while probing at the same time 
the effects of the environment.

Searches for planets around dwarf stars in clusters have not yet 
provided even bona-fide planetary candidates. This result may still be due
to statistical problems, since in particular open clusters are 
tipically loosely populated and the most widely used technique
to detect planets in such environments is the transit method. However,
the increasingly larger number of surveys and the lack of detections
could start to reveal some fundamental differences between planet
formation processes around field and cluster's dwarf stars. It is then 
of primary importance to continue monitoring clusters and 
estabilish on a firmly observational basis if there is indeed a difference
between the planet frequency in these environments and in the field.

In this work we present some preliminary results regarding our survey
toward \object{NGC~6253}. In August 2008, we followed-up with VLT 
three candidate transiting planets we found 
in the region of this cluster. We will dedicate a forthcoming contribution
to the study of planet frequency in the field and in the cluster.
Our observational strategy was tied also 
to the study of cluster's members and surrounding field stars, 
and here we present the results of the complete multi-epoch radial velocity 
campaign toward \object{NGC 6253}. We used FLAMES in MEDUSA mode, 
targeting a total of 204 stars in the region of the cluster. 

In this work we then explore the possibility to search for planetary
companions around dwarf stars of old open clusters by means of the 
radial velocity technique. This detection method would greatly
increase our chances to detect planetary mass companions, allowing us to
overcome the problem of the small sample of stars available in clusters.
However such technique has been applyied systematically only 
to the Hyades cluster so far (Cochran et al. 2002, Paulson et al. 2002,
Paulson et al. 2003)
given the closeness and consequently brightness of its members. In young clusters 
planet detection is significantly hampered and complicated
by stellar activity (Paulson et al. 2004a, Paulson et al. 2004b).
Old open clusters should be better targets, although the faintness of their
members seems to significantly limit the application of the radial velocity 
technique, requiring in general a very large investment of observing time. Here
we discuss an optimized observing strategy aimed at isolating only 
the most promising objects to follow-up with high precision spectroscopy, 
involving the use of photometry, astrometry and 
multi-object spectroscopy. Beside resulting in a great improvement in the 
knowledge of cluster's properties this method minimizes the time needed to 
identify and follow-up best candidates. The application of this technique to the
particular case of \object{NGC~6253} is of special interest given the
characteristics of this cluster.

\object{NGC 6253} has been studied by several authors in the past,
both photometrically (Bragaglia et al. 1997, 
Piatti et al. 1998, Sagar, Munari, \& de Boer 2001, 
Twarog, Anthony-Twarog \& De Lee 2003, Anthony-Twarog, Twarog, \& Mayer 2007,
Montalto et al. 2009, De Marchi et al. 2009), and
spectroscopically (Carretta et al. 2000, Carretta, Bragaglia
\& Gratton 2007, Sestito et al. 2007). In addition,
in Montalto et al.~(2009) we calculated proper motion membership
probabilities.

These studies have demonstrated that \object{NGC 6253} is an
old ($\sim$3.5 Gyr, e. g. Montalto et al.~2009), and metal-rich 
cluster ([Fe/H]=+0.39$\pm$0.07 Sestito et al.~2007; 
[Fe/H]=+0.46 Carretta, Bragaglia \& Gratton~2007),
being in fact one of the most metal-rich open 
clusters of the Galaxy. It is also one of the
few old open clusters located inward the solar ring
(Carraro et al. 2005a, 2005b),
at a Galactocentric distance of around 6 kpc, 
where more prohibitive environmental conditions in general
prevent clusters' survival (Wielen 1971).
\object{NGC 6253} it is also important in the more general context
of stellar population studies, offering an homogeneous
sample of coeval metal-rich stars against which stellar models
at this extreme metallicity can be tested and compared. 
Its peculiar location in the Galactic disk gives the opportunity
to extend toward the inner regions of the Galaxy
the baseline for the study of the Galactic disk
radial abundance, which is a basic ingredient of
Galactic chemical evolution models (Tosi 1996).

This paper is organized as follows: in Sect.~\ref{s:observations}, 
we describe our observations; in Sect.~\ref{s:reduction}, we give a detailed 
description of reductions and calibrations; in Sect.~\ref{s:results}, 
we discuss cluster's membership and binarity in our sample; in Sect.~\ref{s:transits}, we analyze our three transiting
planetary candidates; in Sect.~\ref{s:EA}, we focus our attention on four detached
eclipsing binary systems; in Sect.~\ref{s:EB}, we discuss a new eclipsing binary system
located at the sub-giant branch of \object{NGC~6253}; 
in Sect.~\ref{s:RV_search}, we present an optimized strategy to search for sub-stellar objects
with the radial velocity technique around turn-off stars of old
open clusters; 
in Sect.~\ref{s:simulation}, we discuss a method to
constrain the minimum mass and period of cluster's close binary systems,
detected by RV surveys;  
finally in Sect.~\ref{s:conclusions}, we summarize and conclude.

\section{Observations}
\label{s:observations}
The observations were obtained using the FLAMES facility (Pasquini et al.~2002)
at the UT2 (Kueyen telescope), in Paranal, Chile. FLAMES is the multi-object, intermediate and high resolution 
spectrograph of the VLT. It can access targets over a 25 arcmin diameter field of view,
and it feeds two different spectrographs: UVES and GIRAFFE. While UVES provides the
maximum resolution (R=47000), but can access up to 8 targets at the time, GIRAFFE
has an intermediate resolution (either R$\simeq$25000 or R$\simeq$10000), allowing to target
up to 130 objects at the time or to do integral field spectroscopy. 

As reported in the
previous section, our main purpose was to follow-up three planetary transiting candidates
we found in the field of \object{NGC 6253}. Then, we used UVES in fiber mode, with the
standard setup centered at 580nm. To maximize the scientific output
of our research, we used simultaneously UVES with GIRAFFE, with the high resolution
grating HR9B, covering 21.3nm centered at 525.8nm with a resolution R=25000.

Given the spatial distribution of the transiting candidates we decided to prepare
two different configurations named NGC6253\_A and NGC6253\_B\footnote{This is 
also the nomenclature adopted for the data stored in the ESO archive,
where we further distinguished the different observing nights, e.g. for the first configuration
and the first night we used NGC6253\_A\_N1.}.

\begin{figure}
\center
\includegraphics[width=8cm]{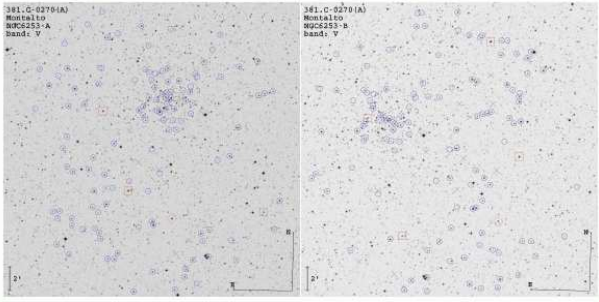}
    \caption{The two UVES/MEDUSA configurations reported in the text. Circles denote the positions
             of the UVES/GIRAFFE targets. The rhombus corresponds to the VLT guide star, and the boxes
             to the three reference stars used by the Field Acquisition Coherent Bundle (FACB) guide
             fibers.
        }
\label{fig:configurations}
\end{figure}

In each configuration the 8 fibers of UVES were allocated in the following manner: 
one fiber was dedicated to the simultaneous calibration, two fibers were allocated
to two transiting candidate host stars, four (three in the second
configuration) fibers to some known cluster's member stars, and one (two in the 
second configuration) to the sky. Since our faintest candidate planet host star
($V=18.247$) was located in between the other two, we were able to observe it in both
configurations in order to improve the $S/N$.

The GIRAFFE fibers were used to target other cluster's members, known photometric variables 
(De Marchi et al.~2009), allocated to the sky or simultaneous calibration lamps.

The data were obtained during three nights between August 17, 2008 and August 25, 2008. 
The exposure time for each configuration was 1875 sec (15 min overhead), for a total observing
time of $\sim$6 hours. The observations were performed in service mode. Table~\ref{tab:obs} provides 
a summary of our data. 

\begin{table}
\caption{
Journal of observations.
\label{tab:obs}
}

\begin{center}
\begin{tabular}{c c c}
\hline
\hline
Plate name & Date & MJD \\
\hline
NGC6253\_A\_N1 & 2008-08-17 & 54695.04805166 \\
NGC6253\_B\_N1 & 2008-08-17 & 54695.01926262 \\
NGC6253\_A\_N2 & 2008-08-20 & 54698.99723017 \\
NGC6253\_B\_N2 & 2008-08-21 & 54699.02557366 \\ 
NGC6253\_A\_N3 & 2008-08-25 & 54703.00230984 \\
NGC6253\_B\_N3 & 2008-08-25 & 54703.03139454 \\
\hline
\end{tabular}
\end{center}
\end{table}

We targeted a total of 204 stars, 44 of which present in both configurations. 
For a given star, we used always the same fiber (apart from the stars in common
to the two configurations) to minimize systematic effects.
We measured the radial velocity for a total of 139 stars.
We were not able to obtain accurate radial velocities for 65 stars.
 
For 24 objects this was probably due to their faint magnitude, since they had 
$R>16$. For the remaining 41 stars, either to their intrinsic characteristics
(hot temperature, high rotation velocities, etc.) or bad fiber positioning, or 
blends.

Among the stars analyzed, 106 have 3 measurements (present only in one configuration), 
and 33 have 6 measurements (present in both configurations). 
We selected preferentially probable cluster's members, on the basis of 
colors, magnitudes and proper motion membership probabilities (MP). 
These probabilities were calculated in Montalto et al.~(2009), considering
likely cluster's members those stars located within a rectangle region
of dimesion 6.3 x 7.9 arcmin inclusive of the cluster's center (see Montalto 
et al.~2009 for details), with magnitude $V<18$, and that have 
MP$>$90$\%$ at $V=12.5$, down to MP$>$50$\%$ at $V=18$. 
The selected stars are representative of different stellar evolutionary stages, and include
a sample of turn-off, sub-giant branch, red-giant branch, red-clump, and blue straggler
stars. We also considered in our target list 16 stars present in the sample of 
variable stars compiled by De Marchi et al.~(2009). These are: the contact binary
(EW) 30341; the RS Canum Venaticorum variable (RS CVn) 44079;
the detached binary systems (EA) 31195,
24487, 38138, 126376; the rotational variables (RO1, RO2, see De Marchi et al.~2009 for details) 
9832, 6430; the long period variables 173273, 15343$_2$, 40819, 16649, 50025, 4306;
the RR Lyrae (RR) 74452, and the $\delta$ Scuti (DSCT) 8420$_6$. One of these stars,
the RS Canum Venaticorum variable (RS CVn) 44079, is a likely cluster's member. 
Finally, seven stars in the spectroscopic sample of Carretta et al.~(2007) and 
Sestito et al~(2007) were included in our list, further improving 
our chances to detect binarity for these objects. They are:
stars 45410, 45412, 45413, and 45414 from Carretta et al.~(2007), and
stars 45404, 45421, 45474 from Sestito et al.~(2007),
all likely members of \object{NGC 6253}. The numeration of these 
stars is that one presented in the catalog of Montalto et al.~(2009)
or in De Marchi et al.~(2010).
The remaining stars in our list were selected considering either
the need to have a sample of field objects against which
comparing the results obtained for cluster's stars, or technical
constraints like that the need to avoid putting the fiber buttons too 
close to each others.

\section{Data reduction, and radial velocity measurement.}
\label{s:reduction}

Data were reduced using the UVES and GIRAFFE pipelines (Ballester et al. 2000,
Blecha et al. 2000) where raw data were bias-subtracted, flat-field corrected,
extracted using the average extraction method,
and wavelength-calibrated. GIRAFFE spectra were wavelength-calibrated using
both prior and simultaneous calibration-lamp spectra, which assure a final
systematic error in radial velocities lower than 100 m/s (Sommariva et al. 2009). 

In particular, for each plate the wavelength calibration was firstly done
using the next morning ThAr frame. Then a drift correction measured
by the 5 simutaneous calibration lamps was applied.
The drift was of the order of few hundred m/s

Finally, sky subtraction was applied.
For what concerns UVES data, echelle orders were flux-calibrated using the 
master response curve of the instrument. 
Finally, the orders were merged to obtain a 1D spectrum.
Radial velocities were obtained from the IRAF fxcor cross-correlation
subroutine. Stellar spectra were cross-correlated with synthetic templates
calculated by SPECTRUM\footnote{See http://www.phys.appstate.edu/spectrum/spectrum.html for more details.}.
The selection of a proper template for each star was mandatory because
the targets have different spectral types and rotational velocities. 
For this reason we used 3 templates: the first calculated for
the sun and valid for stars with effective temperatures in the range  T$_{\rm eff}=$4000-6500 K;
the second calculated for T$_{\rm eff}$=3500 K, which includes the 
strong molecular bands, and is valid for the
coolest stars in our sample; the last calculated for T$_{\rm eff}$=7000 K,
valid for the hottest stars in our sample. In addition, the template
for hot stars was divided in 7 sub-templates, calculated for rotational
velocities equal to 0, 25, 50, 75, 100, 200, and 300 km/s.
For each star we carefully compared the observed spectrum with our templates
and, according to the visible spectral lines and to a rough estimation of the
rotation, we chose the most appropriate one.

We calculated for each star the mean radial velocity ($\overline{RV_{obs}}$),
and the dispersion ($\sigma_{\rm obs}$):

\begin{equation}
\overline{RV_{obs}}=\sum_{i}\,\frac{RV_{i}}{N}
\end{equation}

\begin{equation}
\sigma_{obs}=\sqrt{\sum_{i}\frac{(RV_{i}-\overline{RV_{obs}})^{2}}{(N-1)}}
\end{equation}

\noindent
where $RV_{i}$ is the i-th radial velocity measurement, and $N$ is the total
number of measurements for each star. 

To determine the threshold for binary detection,
we subdivided the stars in seven magnitude bins 0.5 mag wide between $11<R<14.5$,
plus an additional large magnitude bin between $14\le R<18$ (to account for the fewer number
of stars in this magnitude range). We calculated the mean dispersion in each magnitude bin 
excluding stars with $\sigma_{obs}>1$ km/s ($\sigma_{obs}>2$ km/s, in the fainter large bin),
applied a $\sigma$-clipping algorithm excluding all stars above $3.5\times\sigma$ from
the mean, and calculated the final value of the mean dispersions. We then performed a linear
least square fit of the resulting values separating the brighter bins from the fainter
one. Our result for the best-fit mean dispersion of $constant$ stars ($\overline{\sigma_{\rm obs}}$)
is given by the following equations:

\begin{equation}
\overline{\sigma_{obs}}=0.030\,R\,-0.130\,\,\,\,\,\,R<14.5
\end{equation}

\begin{equation}
\overline{\sigma_{obs}}=0.394\,R\,-5.407\,\,\,\,\,\,R\ge14.5
\end{equation}

\begin{figure}
\center
\includegraphics[width=8cm]{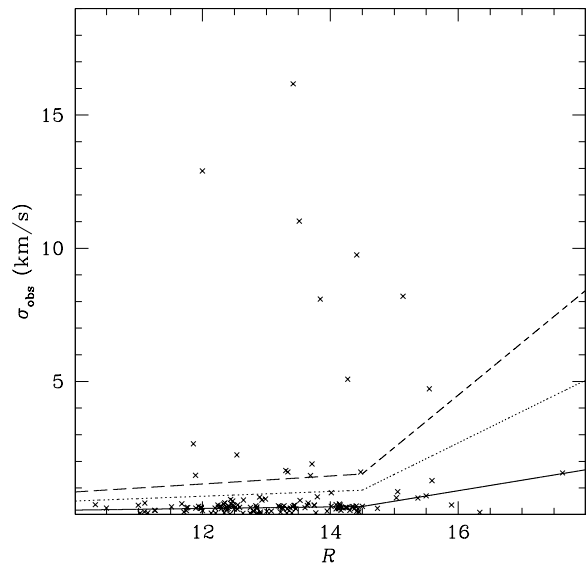}
    \caption{Observed radial velocity dispersions ($\sigma_{\rm obs}$) 
             against $R$-band magnitude. The continuous line indicates our best-fit
             for the typical dispersion ($\overline{\sigma_{obs}}$) of radial velocity 
             $constant$ stars. The dotted and dashed lines are the $3\times\overline{\sigma_{obs}}$,
             and the  $5\times\overline{\sigma_{obs}}$ thresholds. 
        }
\label{fig:rms_r}
\end{figure}

\noindent
We adopted a conservative threshold for binary detection given that
only stars having $\sigma_{\rm obs}>5\times\overline{\sigma_{\rm obs}}$ were considered
candidate spectroscopic binaries. In Fig.~\ref{fig:rms_r}, we show the mean
dispersion as a continuous line, and the $3\times\overline{\sigma_{obs}}$, 
$5\times\overline{\sigma_{obs}}$ as dotted and dashed lines respectively.

Throughout this paper we will call radial velocity variables simply {\it candidate 
close binary systems}, or {\it close binary systems}, 
though it is clear that they could also be multiple systems, or that the radial
velocity variations may have other physical explanations. In particular, we used the 
adjective {\it close}, because our observations span a period of only eight days, and
even when the binarity is inferred by means of a comparison with past literature results,
those observations are separated at most by eight years with respect to our observations
\footnote{This is the case of the observations presented in Carretta et al. (2007).
Sestito et al. (2007) measurements were obtained four years before us.}. 

In Table~2 and in Table~3, we present the measured radial
velocities for the analyzed objects, distinguishing between stars which are likely 
to be proper motion cluster's members (Table~2), and stars that are proper 
motion non members, or with doubtful/absent proper motions (Table~3).

\section{Results}
\label{s:results}

\subsection{Cluster's mean radial velocity}

The recession radial velocity of the cluster was calculated considering 
only proper motion cluster's members. Moreover, we excluded stars which are likely close
binaries (see the previous section). Then we calculated the mean radial velocities
($\overline{RV_{obs}}$) of all the remaining stars, and used them to derive 
the cluster's mean radial velocity ($\overline{RV_{obs}}$ ),
after applying an iterative 3-$\sigma$ clipping algorithm, to exclude also
long period binaries or residual contaminants. The resulting
mean radial velocity of the cluster obtained after this procedure is
$\overline{RV_{cl}}\pm\overline{\sigma_{cl}}=(-29.11\pm0.85)$ km/s, where the error is the
error of the mean, and the root mean square of the residuals is 
$\sigma_{cl}=4.26$ km/s, obtained from the remaining sample of 25 stars.

\begin{figure}
\center
\includegraphics[width=8cm]{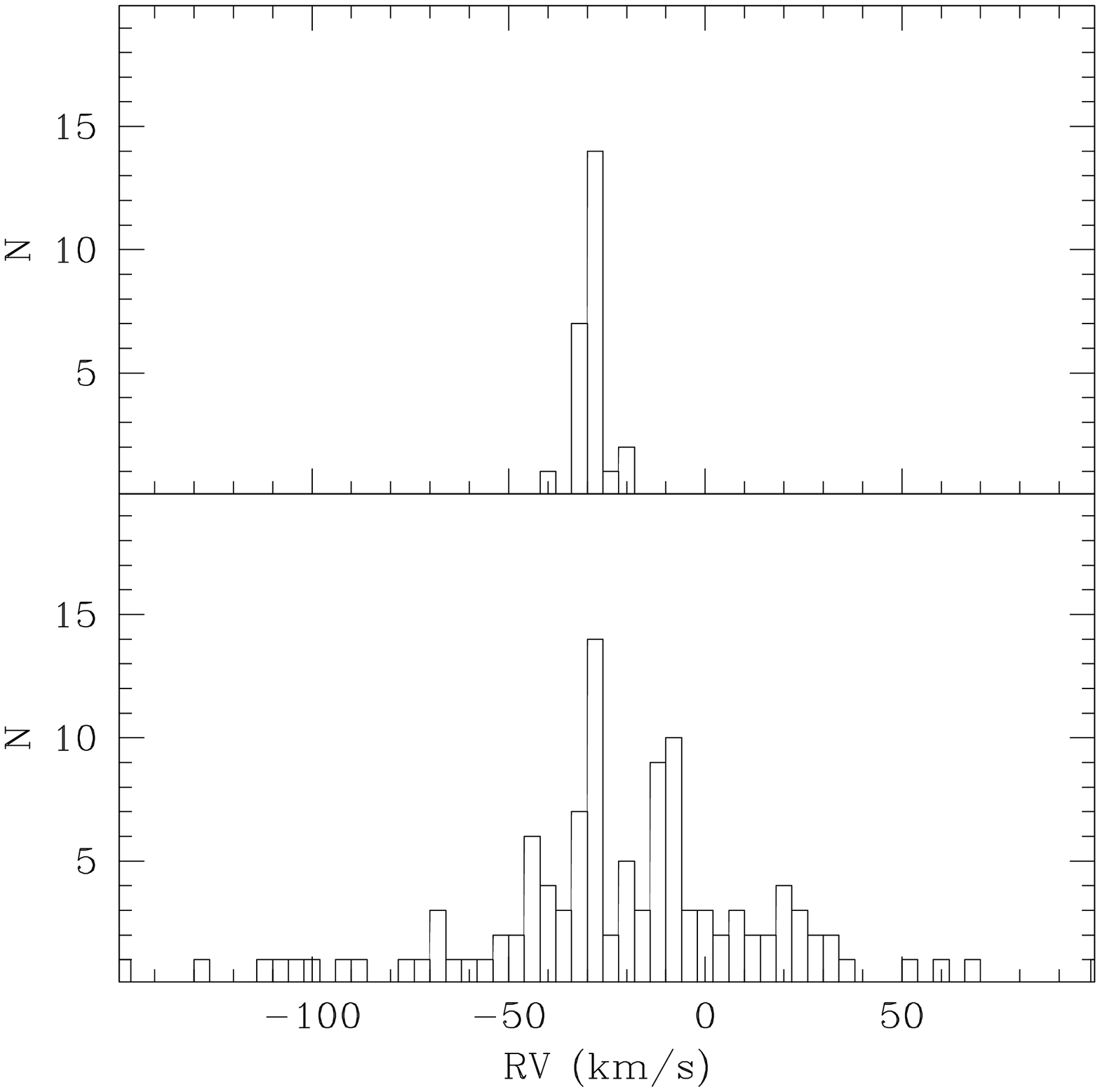}
    \caption{Upper panel: radial velocity distribution of stars considered as cluster's members
             on the basis of proper motions and radial velocities;
             Bottom panel: radial velocity distribution of all the remaining stars in
             the sample (see text).
        }
\label{fig:hist_rv}
\end{figure}

The cluster's velocity dispersion seems quite high, since typical values for open clusters
of similar richness are around 1-2 km/s. This might indicate that there are still long-period
binaries inflating the RV distribution. In Fig.~\ref{fig:hist_rv}, we present the radial velocity
histogram for the 25 stars used above to calculate the cluster's mean radial velocity (upper panel),
and for all the remaining objects (bottom panel), in bins of 4 km/s. Among the stars not considered
in our calculation there are other very likely members, which will be discussed in the next Section.

Considering only stars with radial velocities within 1-$\sigma$ from the mean, the result is
$\overline{RV_{cl}}\pm\overline{\sigma_{cl}}=(-28.98\pm0.12)$ km/s, and the
scatter $\sigma_{cl}=0.4$ km/s, calculated from a sample of 10 stars.
These estimates are in reasonable agreement with previous literature results. Sestito et al.~(2007),
obtained a mean radial velocity equal to (-29.71$\pm$0.79) km/s from a sample
of 4 stars. Carretta et al.~(2007), obtained (-28.26$\pm$0.29) km/s from
a sample of 4 stars.

\subsection{Proper motion and radial velocity members}

Among the sample of proper motion cluster's members,
stars that have mean radial velocities ($\overline{RV_{obs}}$) satisfying the condition:

\begin{equation}
\left |\,\overline{RV_{obs}}-\overline{RV_{cl}}\,\right | <\,3\,\sigma_{cl},
\end{equation}

\noindent
were considered also radial velocity members. 
In Fig.~\ref{fig:rbr_cl}, we present the ($R,B-R$) color magnitude
diagram for proper motion selected cluster's members (small black points),
where stars having also radial velocity measurements are enlighted
by colored squares. Moreover, stars that are likely close binaries 
(as obtained from our multi-epoch radial velocities), are framed in black squares. 
We overplot to the figure the best-fit isochrone taken from the Padova 
database (age=3.5 Gyr, Z=0.03, continuous line), and also the equal mass 
binary sequence (dashed line). The red squares represent stars that met
both our proper motion, and radial velocity membership criteria. 
They are in total 30 stars. As shown in Fig.~\ref{fig:rbr_cl}, 
these stars populate the turn-off, sub-giant, red-giant branch, red-clump and
blue straggler regions of the cluster, and are considered secure cluster members. 
Among them, the following objects are particularly interesting:

\begin{figure*}
\center
\includegraphics[width=17cm]{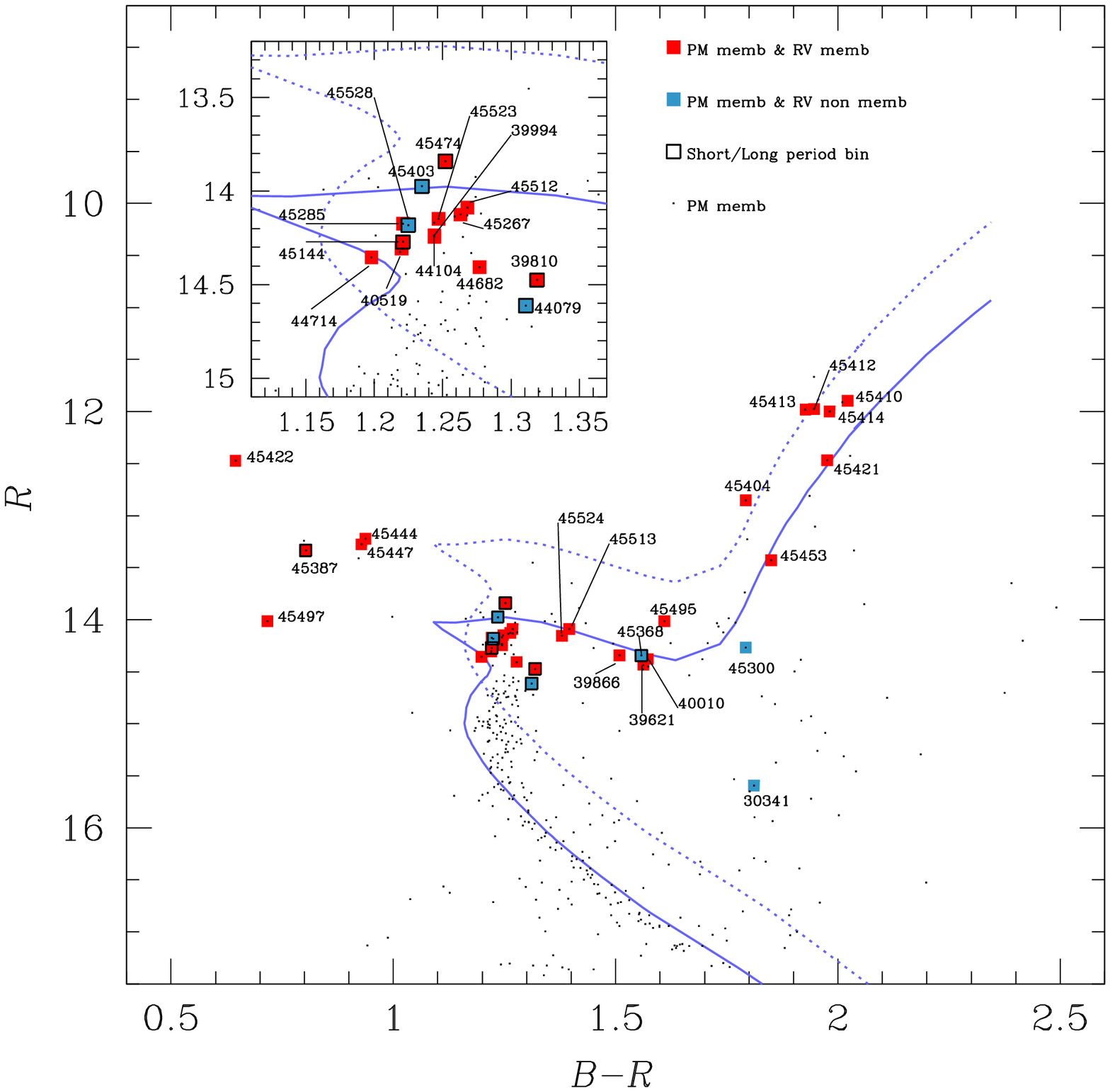}
    \caption{$R$-band magnitude, and $B-R$ color, for proper motion 
     selected members (small black points). Red squares indicate stars that 
     are also radial velocity members, steel blue squares stars that appear radial 
     velocity non members, and black framed squares stars that are likely close binary systems.
        }
\label{fig:rbr_cl}
\end{figure*}

\begin{figure*}
\center
\includegraphics[width=17cm]{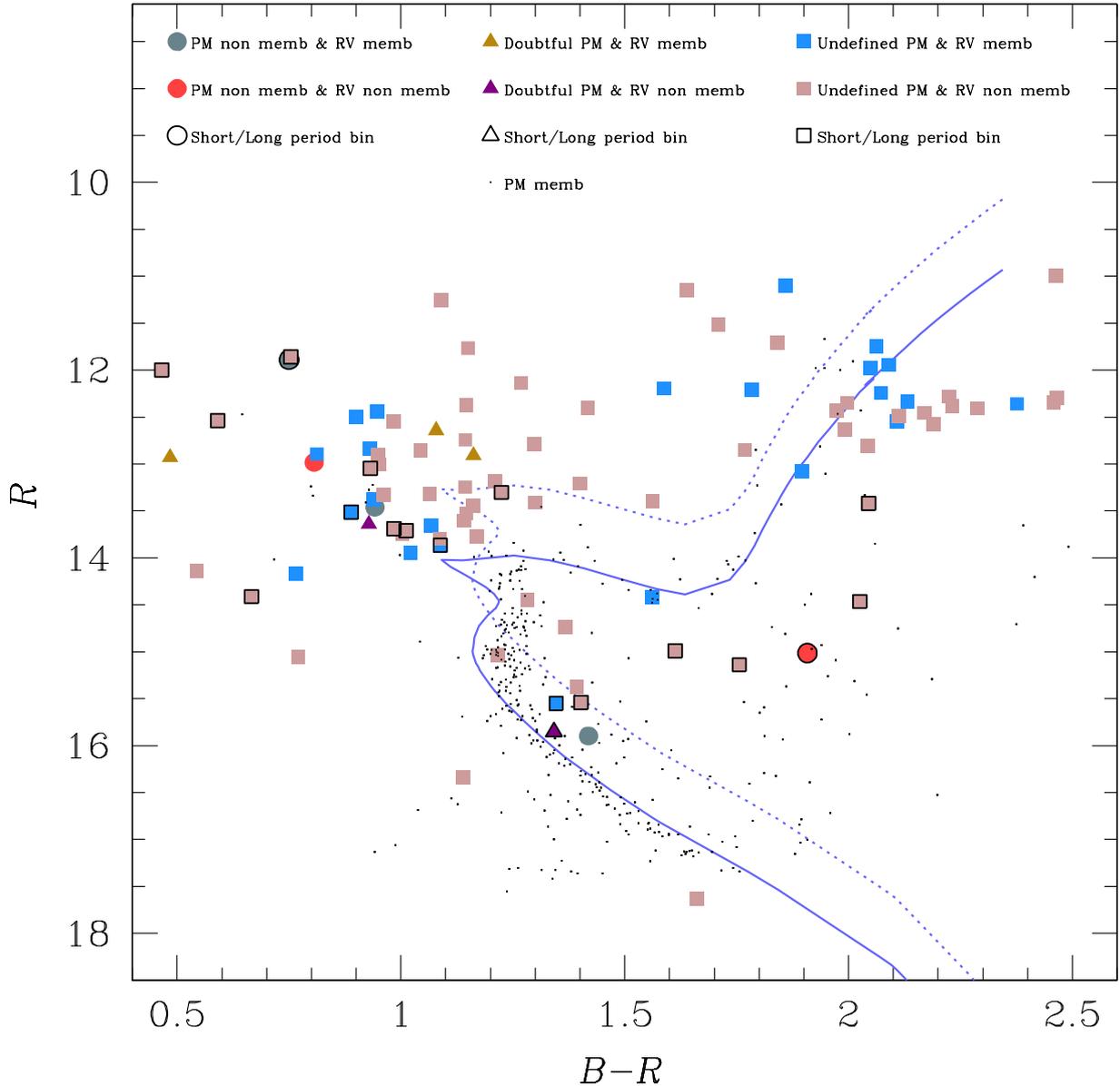}
    \caption{Same as Fig.~\ref{fig:rbr_cl}, but for proper motion non members or for stars 
     which proper motion membership probabilities are either doubtful or absent, as explained
     in the legend.
        }
\label{fig:rbr_nm}
\end{figure*}

\subsection{Star 39810}

Star 39810 has a high proper motion membership probability (MP=95\%), its mean radial velocity
agrees with that one of the cluster ($\overline{RV}_{obs}$-$\overline{RV_{cl}}=$0.037$\sigma_{cl}=0.159$km/s), and
it is located in the turn-off region of the cluster. 
Since from our repeated radial velocities we measured a dispersion $\sigma_{obs}=$1.595 km/s, 
($\sigma_{obs}/\overline{\sigma_{obs}}=5.242$), we conclude that this object is also a 
candidate cluster's close binary system. This object is discussed in more detail in
Sect.~\ref{s:simulation}.

\subsection{Star 45387}
Star 45387 is a proper motion and radial velocity cluster's member 
(MP=97$\%$, $\overline{RV}_{obs}$-$\overline{RV_{cl}}=$0.115$\sigma_{cl}=0.492$ km/s) located in the blue straggler region,
and a likely close binary system ($\sigma_{obs}=$ 1.602 km/s, $\sigma_{obs}/\overline{\sigma_{obs}}=5.932$).

\subsection{Star 45144}

Star 45144 is a proper motion and radial velocity cluster's member 
(MP=97$\%$, $\overline{RV}_{obs}$-$\overline{RV_{cl}}=$1.380$\sigma_{cl}=5.879$ km/s) in the turn-off region.
It is a likely close binary system ($\sigma_{obs}=5.081$ km/s, $\sigma_{obs}/\overline{\sigma_{obs}}=17.041$).

\subsection{Star 45474}

Star 45474 is a proper motion and radial velocity cluster's member 
(MP=97$\%$, $\overline{RV}_{obs}$-$\overline{RV_{cl}}=$1.712$\sigma_{cl}=7.294$ km/s) in the turn-off region,
and a likely close binary system ($\sigma_{obs}=$ 8.090 km/s, $\sigma_{obs}/\overline{\sigma_{obs}}=28.360$).
This star was observed also by Sestito et al.~(2007), that determined a radial velocity 
equal to $(33.49\pm2.87)$ km/s
with a difference of 11.67 km/s with respect to our own mean radial velocity, confirming the binarity of the object.

\subsection{Star 45413}
Star 45413 at the red-clump, proper motion and radial velocity member
(MP=91$\%$, $\overline{RV}_{obs}$-$\overline{RV_{cl}}=$0.884$\sigma_{cl}=3.767$ km/s),
does not present indication of binarity from our dataset
($\sigma_{obs}=$0.467 km/s, $\sigma_{obs}/\overline{\sigma_{obs}}=$1.129),  though a 
comparison with the measurement of Carretta et al.~(2007) gives a difference of 12.207 km/s
with respect to our mean radial velocity, suggesting that the star is a 
binary system.

\subsection{Star 44682}

Star 44682 at the turn-off, is a double-lined 
spectroscopic binary, probably with a period much longer than our
observing window, since from our radial velocities this star is
not variable
(MP=95$\%$; $\overline{RV_{obs}}$-$\overline{RV_{cl}}=$2.774$\sigma_{cl}=11.819$ km/s,
$\sigma_{obs}=$ 0.326 km/s, $\sigma_{obs}/\overline{\sigma_{obs}}=1.079$).

\subsection{Star 45421}

Star 45421 at the red-giant branch,
was already classified as a potential binary star by Sestito et al.~(2007).
Our estimated mean radial velocity differs from their measurement
by $\sim2.7$ km/s. It appears that this star is a likely cluster's close
binary, though from our measurements it was not classified as
a radial velocity variable star (MP=94$\%$; $\overline{RV_{obs}}$-$\overline{RV_{cl}}=$2.519$\sigma_{cl}=$10.730 km/s,
$\sigma_{obs}=$0.339 km/s, $\sigma_{obs}/\overline{\sigma_{obs}}=$1.389).

\subsection{Stars 45404, 45412, 45414, 45410}

These stars were observed also by Sestito et al.~(2007, star 45404), 
and Carretta et al.~(2009, stars 45412, 45414, 45410\footnote{Star 45410 
in common with Carretta et al.~(2007), it is 
considered a proper motion cluster's member, although its position 
in the sky is slightly outside 
the region where we considered reliable our proper motions (Montalto et al.~2009).
}). They are not RV variables, and our mean radial velocities agree with
the measurements of those authors (within their uncertainties), and 
with our radial velocity cluster's membership criterium. 

\subsection{Proper motion members and radial velocity non members}

Stars indicated by steel blue squares (Fig.~\ref{fig:rbr_cl})
have mean radial velocities that do not satisfy our 
cluster's membership radial velocity criterium. They are in total six stars.
Five of them are likely cluster's close binaries.

\subsection{Star 44079}

Star 44079, at the cluster's turn-off, 
(MP=90$\%$; $\overline{RV_{obs}}$-$\overline{RV_{cl}}=$5.392$\sigma_{cl}=22.968$ km/s,
$\sigma_{obs}=$35.258 km/s, $\sigma_{obs}/\overline{\sigma_{obs}}=100.587$)
is a spectroscopic binary system.
De Marchi et al.~(2009) report that this star is a RS CVs 
star with period $\sim$2.18 days, 
and presents signs of activity and also a shallow eclipse ($\sim0.02$ mag).

\subsection{Star 45368}

Star 45368, located at the sub-giant branch is an
eclipsing binary double-lined system, as we determined by visual inspection of the 
light curve\footnote{This object is not included in the list of variable
stars of De Marchi et al.~(2009)}. The binarity is also confirmed
by our radial velocity measurements 
(MP=95$\%$; $\overline{RV_{obs}}$-$\overline{RV_{cl}}=$6.146$\sigma_{cl}=$26.183 km/s,
$\sigma_{obs}=$109.131 km/s, $\sigma_{obs}/\overline{\sigma_{obs}}=363.682$).
It will be discussed in more detail in Sect.~\ref{s:EB}.

\subsection{Stars 45528}

Star 45528 at the turn-off is a double-lined spectroscopic binary system
(MP=97$\%$; $\overline{RV_{obs}}$-$\overline{RV_{cl}}=$14.690$\sigma_{cl}=$62.581 km/s,
$\sigma_{obs}=$20.931 km/s, $\sigma_{obs}/\overline{\sigma_{obs}}=70.842$).

\subsection{Stars 45403}
Star 45403 at the turn-off is a double-lined spectroscopic binary system
(MP=97$\%$; $\overline{RV_{obs}}$-$\overline{RV_{cl}}=$5.177$\sigma_{cl}=$22.054 km/s,
$\sigma_{obs}=$39.078 km/s, $\sigma_{obs}/\overline{\sigma_{obs}}=135.101$).

\subsection{Star 45300}

Star 45300 is located close to the red-giant branch of the cluster.
Its deviant radial velocity would suggest that this object could be a 
candidate cluster's close binary,
although we have no possibility to check this hypothesis. From our
radial velocities the object is not variable
(MP=94$\%$; $\overline{RV_{obs}}$-$\overline{RV_{cl}}=$12.564$\sigma_{cl}=$53.524 km/s,
$\sigma_{obs}=$0.256 km/s, $\sigma_{obs}/\overline{\sigma_{obs}}=0.859$).

\subsection{Star 30341}

Star 30341 is instead 
almost certainly a field contaminant with similar proper motion
of cluster's stars, as indicated by its deviant radial velocity 
(MP=87$\%$; $\overline{RV_{obs}}$-$\overline{RV_{cl}}=$10.100$\sigma_{cl}=$43.025 km/s,
$\sigma_{obs}=$1.277 km/s, $\sigma_{obs}/\overline{\sigma_{obs}}=1.734$),
and position in the CMD. Note that its proper motion membership probability
is the lowest among the sample of proper motion cluster's members listed
in Table~3. De Marchi et al.~(2009), classified this star
as an eclipsing contact binary system (EW) with period 0.27 days,
and semi-amplitude $\sim0.02$ mag. From our radial velocities, 
the object is not variable.
Given the small dispersion of the radial velocity measurements, and the
very small period determined photometrically, we argue that this object may
be a rotational variable, rather than a contact binary. As stated 
in De Marchi et al.~(2009), it is very difficult to distinguish between
these two classes of variables from the photometry alone.

In summary, given the high proper motion membership probabilities of these stars, 
and their position in the CMD, it is likely to believe that their are all likely cluster's
members (with the exception of star 30341). In particular, 
their deviant radial velocities could be explained by
the large dispersions we observed for stars 44079, 45368, 45528, 45403, and by
long periods and massive companions for stars 44682, 45421, and 45300, 

\subsection{Proper motion non members or stars with doubtful/undefined proper motions}

In Fig.~\ref{fig:rbr_nm}, we present the results for stars
which are proper motion non members, or have doubtful/undefined proper
motions, as indicated by the different symbols (see the legend).
Different colors separate radial velocity members from non members.
Framed symbols highlight close binaries among this sample. None of the few stars
with proper motions met our selection criteria for cluster's membership,
and most stars have undefined proper motions. 

The median radial velocity of this sample of stars
is RV$_{field}=-8.98$ km/s (excluding close binaries)
and the root mean square of the residuals is $\sigma_{field}=48.40$ km/s. Fitting a 
Gaussian function to this distribution we expect $(9\,\pm\,3)$ stars 
with radial velocity consistent with cluster's membership
($\overline{RV_{cl}}=-29.11\pm$ km/s, 12.78 km/s$=3\sigma_{cl}$).
We count instead 14 potential radial velocity members. 
Looking at Table~\ref{tab:rv_fld}, considering the radial distances of these objects from 
the cluster's center, and their positions in the color magnitude diagram, the
most likely cluster's members are: the detached binary (EA) 31195, located
in between the cluster's main sequence and the cluster's equal mass binary sequence
(Fig.~\ref{fig:rbr_nm}); the sub-giant branch star 25450$_{2}$; the blue straggler
candidates 45392, 45396, 45409, 45427, and 45433. These candidate members must be considered
with caution, since their proper motion membership probabilities are not very large ($\le85\%$).

\subsection{The frequency of binary systems}

The frequency of binaries among cluster's and field stars can be investigated thanks
to photometric (Montalto et al. 2009, De Marchi et al. 2009) 
and radial velocity results (Sestito et al.~2007, Carretta et al.~2007 and this work). 
We consider a star as a likely binary system if either
the photometry\footnote{We did not considered in the binary list pulsational variables like 
RR Lyrae, and $\delta$ Scuti stars.}
or the spectroscopy allowed us to classify the star as a binary.

Among our sample of 35 likely cluster's members, we found that 12 objects are 
likely cluster's binary systems giving a binary frequency of f$_{bin}=$(34$\pm$10)$\%$, 
where the error accounts for Poisson statistic\footnote{Star 30341 is considered a field star, 
and star 45300 is considered a long period cluster's binary.}. 
The frequency remains essentially inaltered including also the sample of 7 stars indicated
in the previous paragraph as likely members (among which two are variables), f$_{bin}=$(33$\pm$9)$\%$. 
This result is in good agreement with previous estimates (e. g. Bragaglia et al., 1997; 
Montalto et al., 2009), though it is certainly underestimating the real value, 
since in general we are not complete for long period binary systems 
and very small mass companions. 

However, an homogeneous comparison with the field's binary
frequency can be done only for binaries detected from our own surveys 
(both photometric and spectroscopic), since no field objects were spectroscopically
observed by other authors in the past years. Just to 
distinguish the frequency of these binaries from the previous estimate, 
we use here the term {\it frequency of close binaries} (f$_{cl}$), 
because both our photometric and spectroscopic surveys covered a few nights of observations.

Then 10 stars out of 35 are likely cluster's close binary systems
of \object{NGC 6253}, which gives:
f$_{cl}=$(29$\pm$9)$\%$ (or 29$\%$$\pm$8$\%$ considering the seven stars in the
previous paragraph) for the cluster, and equal
to f$_{cl,field}=$(22$\pm$5)$\%$ (excluding the seven stars of the previous
paragraph) for the field. Our estimated cluster's binary frequency appears then 
higher than the field binary frequency, although they are consistent within
the uncertainties. We note also that the sample of cluster's stars analyzed is still small, and there
may be some selection effects. De Marchi et al.~(2009) 
observed that the class of main-sequence rotational variables is the most
numerous, as observed in the surrounding field of \object{NGC 6253}. 
In our sample, there are in fact only two rotational variables.
More in general, in this work we focused
on the binarity of turn-off stars, and evolved stars, whereas the study
of De Marchi et al.~(2009) analyzed the variability of fainter objects.

Mermilliod et al.~(2009) obtained an overall binary frequency equal to 30$\%$
from the analysis of 13 nearby open clusters, considering their 19-years CORAVEL survey,
which would be consistent with our estimates.

Mathiew R. D. et al.~(1990) derived a frequency of binaries with periods less
than 1000 days comprised between 9$\%$ and 15$\%$ among the solar mass M67 members.  
Latham et al.~(2002) obtained a frequency equal to $(15.8\pm1.5)\,\%$ for the halo population
finding no obvious difference with the disk populations. An accurate comparison of our
results with previous literature findings is beyond the purpose of this work, however we note
that our estimated frequency of close binary systems ($f_{cl}$) appears somehow higher than what
derived by these authors in other environments, even if the difference is only significant 
at $\sim$1.6-$\sigma$, considering our errors.

\section{Transiting planetary candidates}
\label{s:transits}

\begin{figure}
\center
\includegraphics[width=8cm]{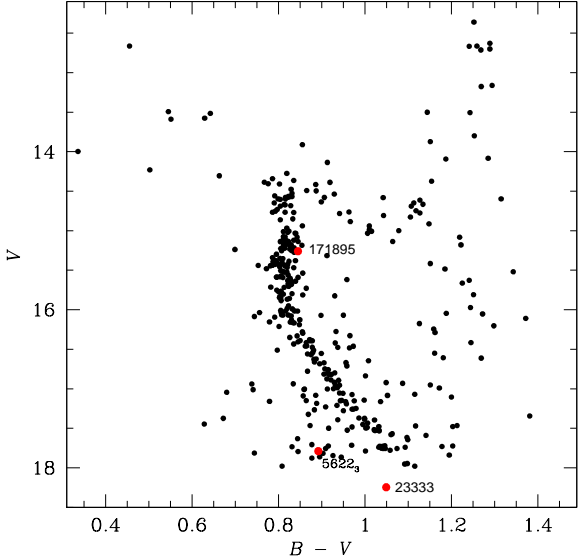}
    \caption{$V$, $B-V$ color magnitude diagram of proper motion cluster's members of \object{NGC 6253}, 
        with highlighted the positions of the three planetary transiting candidates discussed in the text.
        }
\label{fig:CMD_cand}
\end{figure}

In this Section, we present a preliminar analysis of the three planetary transiting candidates we 
found in the field of \object{NGC~6253}, based on the photometry data acquired at the 
La Silla 2.2m Telescope, and on the UVES spectroscopy. In Fig.~\ref{fig:CMD_cand}, we
present the color magnitude diagram of proper motion cluster's members together with the colors and 
magnitudes of the transiting candidates. None of these candidates is located in the region
were our proper motions are reliable. Moreover, from Fig.~\ref{fig:CMD_cand}, it appears
that only star 171895 may be compatible with cluster's membership, since it is located
in the turn-off region. The other objects are likely field stars.

\subsection{Star 171895}

\noindent
The magnitude of the target star is $V=15.260\pm0.002$. From the analysis of the UVES spectra,
we obtained the following parameters:
 $\rm T_{eff}=5720\pm50 K$, $\rm log(g)=4.50\pm0.20$, $\rm [Fe/H]=+0.36\pm0.02$. Using these
parameters and uncertainties and the isochrones with metallicity $\rm Z=0.03$
with ages comprised between $\rm log(age[Gy])=7.8$ and $\rm log(age[Gy])=10.25$ 
taken from the Padova database we constrained the mass, radius
and age of the star obtaining: $\rm M=(1.07\pm0.04)M_{\odot}$, 
$\rm R=(1.04\pm0.07)R_{\odot}$ and $\rm age=(8.9\pm0.7)\,Gyr$. 

\begin{figure*}[!]
\center
\includegraphics[width=15cm]{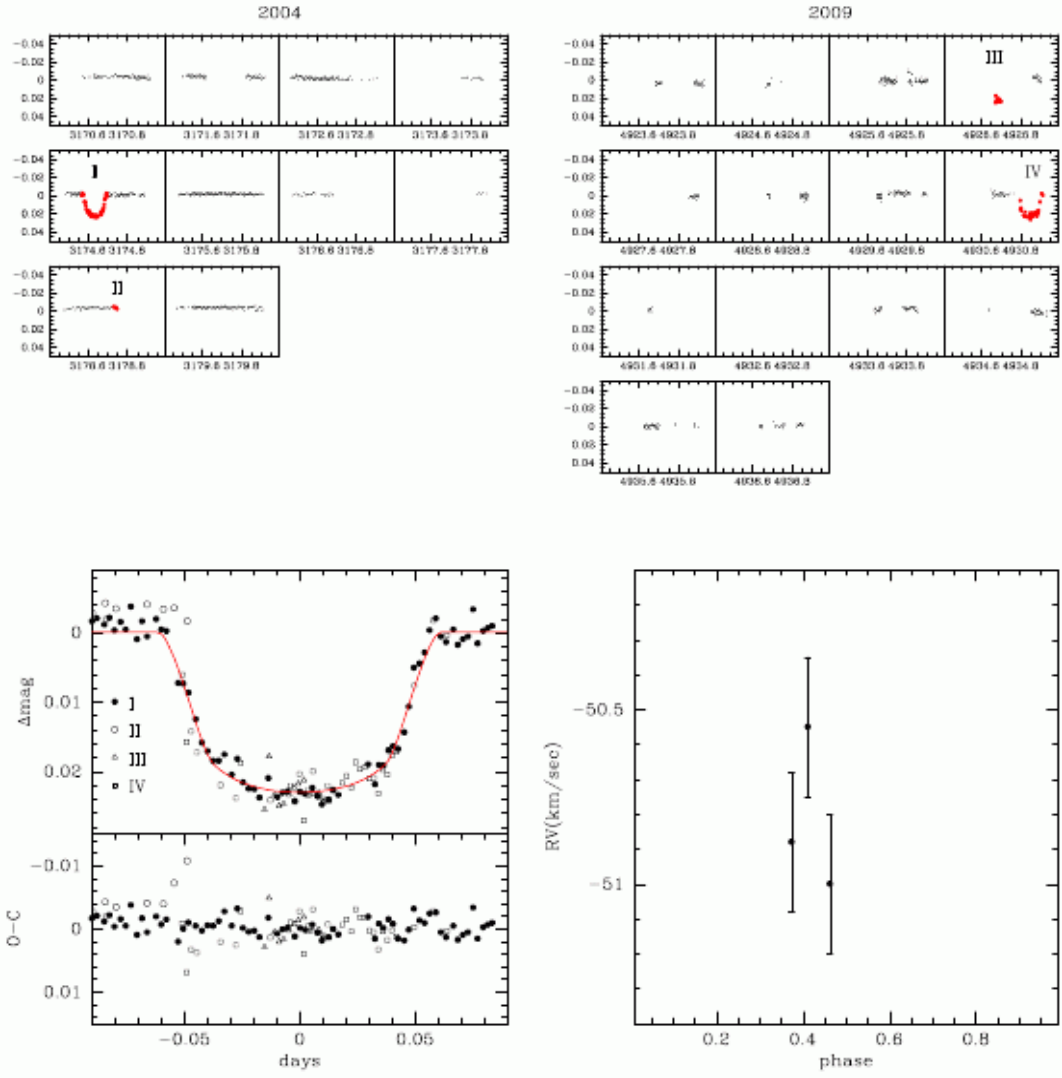}
    \caption{{\it Upper left:} lightcurve of star 171895 relative to the
    2004 observing season. Red points highlight transit epochs, also indicated
    by the roman numeration. {\it Upper right:} lightcurve of star 171895 relative
    to the 2009 observing season. {\it Lower left, top panel:} folded lightcurve. The continuous line
    indicates the best-fit model obtained using the Mandel \& Algol~(2002) algorithm,
    considering as properties of the host star those derived from the UVES
    spectroscopy. The roman numeration and the associated symbols are relative to the
    transit epochs shown in the upper panels. {\it Lower left, bottom panel:} observed minus
    model residuals. {\it Lower right:} UVES spectroscopic measurements.
        }
\label{fig:chip8_cand}
\end{figure*}

\noindent
In Fig.~\ref{fig:chip8_cand} (upper panels) we present the entire lightcurve of the object
obtained using the WFI data. The photometry of the 2009 observing season was more noisy than 
the photometry of 2004, since we observed in April (at tipically higher airmasses than
in June 2004), and during bright time (dark time in 2004). However,
we detected three evident transit events (denoted by the roman numbers I, III, and IV in
Fig.~\ref{fig:chip8_cand}) of around 0.024 mag depth. The first (full) transit
was detected in 2004, the other two (a partial transit during the flat bottom region
and a full transit) in 2009
\footnote{The 2009 photometric observations will be accurately described in a forthcoming
paper, they have been obtained thanks to Max Planck Institute for Extraterrestrial Physics 
reserved time.}. The period of the transiting object was deduced using at first
the two transits of 2009, which allowed us to guess the closest integer period
to the real period (that is four days), and then folding the lightcurve finding
the best solution consistent with all our photometric measurements and able to overlap the 
two full transits. We assumed a $constant$ period. The result is $\rm P=4.16164$ days. 
Submultiples of this period are then excluded by our photometry. On the basis of this
procedure we also deduced that a few photometric measurements acquired just at the end of the
nineth night in 2004 should have been located just at the ingress of the transit 
(see the epoch denoted by the roman number II). In the folded lightcurve presented in Fig.~\ref{fig:chip8_cand}
(bottom left panel) we note that the points acquired during that night (open circles) 
present a slight photometric offset with respect to the other 
measurements ($\sim0.002$ mag), and that the two points  
just inside the transit present relatively large residuals with
respect to the best transit model matching the observations denoted
by the red, even accounting for the photometric offset.
Wether this is an indication that assuming a $constant$ period is
not correct it is not clear from these observations.
A few other measurements acquired during epoch IV (open boxes) and with similar phases
$\sim-0.05$ days in the same Figure, appear to have large residuals as well,
however this is more likely due to the lower photometric quality of the
2009 observing season. Other photometric measurements are cleary necessary 
to clearly understand this system.

The best-fit model was obtained using the 
Mandel \& Agol (2002) algorithm, and a quadratic limb darkening law, 
where the limb darkening coefficients were selected
from the table of Claret (2000) accordingly to the spectral type of the host
star. The stellar parameters (mass and radius) were also fixed to the 
mean values we obtained from spectroscopy. The model well reproduces
the shape of the observed transits. The RMS of the
fit is 0.002 mag, the radius of the transiting object we obtained is 
$\rm R_{pl}\,=\,1.49\,R_{jup}$, and the inclination $87^{\circ}$.

The UVES radial velocity measurements are shown in Fig.~\ref{fig:chip8_cand} 
(lower right panel), and presented in Table~\ref{tab:cand_171895}. Since we did not 
know the orbital period at the time of the
UVES observations (only one transit was detected in 2004, see above), it was not
possible to accurately plan the spectroscopic follow-up of this object. The
UVES measurements were unfortunately acquired at very similar orbital phases, 
as shown in Fig.~\ref{fig:chip8_cand} (lower right panel). As a consequence
it is not possible to derive an orbital solution. The measurements are 
compatible within their errors, which is 200 m/s. Observations at other orbital phases
and of higher precision are necessary to accurately constrain the mass of the transiting
object. The epoch ($E$) of the eclipses is given by:

\begin{displaymath}
(HJD - 450000.)\,=\,3162.15268\,+\,4.16164\,\times\,E.
\end{displaymath}

\begin{table}
\caption{
Journal of UVES data for star 171895.
\label{tab:cand_171895}
}

\begin{center}
\begin{tabular}{c c c}
\hline
\hline
Epoch & HRV(km/sec) & $err_{HRV}$(km/sec) \\
\hline
   2454695.53216 &   -51.00 & 0.2\\
   2454699.53815 &   -50.55 & 0.2\\
   2454703.54365 &   -50.88 & 0.2\\
\hline
\end{tabular}
\end{center}
\end{table}

\subsubsection{Cluster's membership}

\begin{figure}
\center
\includegraphics[width=8cm]{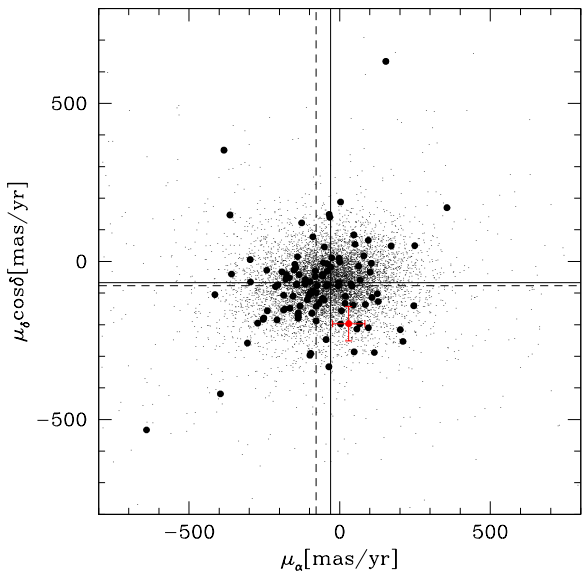}
    \caption{UCAC2 proper motion diagram. Big black points denote likely
    cluster's members (see text). Small black dots stars with radial distance from
    the cluster's center comprised between $10^{'}$ and $40^{'}$, and the 
    red dot the proper motion of star 171895. The dashed lines indicate
    the median proper motion of cluster's stars, and the continuous lines
    the median proper motion of field stars.
        }
\label{fig:pm_cand}
\end{figure}

As demonstrated above, star 171895 is a very metal rich star, and it appears
located in the cluster's turn-off region.
Since the target star has no proper motion in our own catalog (see above), we retrieved 
proper motions from the UCAC2 catalog. In particular we cross-matched
our catalog with UCAC2, isolating likely cluster's members on the
basis of our proper motions. In Fig.~\ref{fig:pm_cand}, we show the proper motion of the
target star (red point), together with the (UCAC2) proper motion of likely cluster's
members (big black dots), and with those of likely field stars with radial distance
from the cluster's center comprised between 10$^{'}$ and 40$^{'}$. The median values
and the  standard deviations of the right ascension and declination proper motion distributions are 
$\overline{\mu_{\alpha}}=-78$ mas/yr, $\overline{\mu_{\delta\,cos(\delta)}}=-77$ mas/yr, 
$\sigma_{\mu_{\alpha}}=153$ mas/yr,  $\sigma_{\mu_{\delta\,cos(\delta)}}=141$ mas/yr, for likely cluster's members, and 
$\overline{\mu_{\alpha}}=-30$ mas/yr, $\overline{\mu_{\delta\,cos(\delta)}}=-67$ mas/yr, $\sigma_{\mu_{\alpha}}=179$ mas/yr,
 $\sigma_{PM_{\delta\,cos(\delta)}}=116$ mas/yr for likely field stars. The proper motion of the target star is 
$\mu_{\alpha}=30$ mas/yr, $\mu_{\delta\,cos(\delta)}=-197$ mas/yr. From these numbers it appears that UCAC2 proper motions
are not very effective in distinguishing among field and cluster's stars given the large
errors. The proper motion of the target star is consistent 
within 0.7$\sigma_{\mu_{\alpha}}$, and 0.85$\sigma_{\mu_{\delta\,cos(\delta)}}$ with
the likely cluster's members distribution and within 0.3$\sigma_{\mu_{\alpha}}$, and 1.12$\sigma_{\mu_{\delta\,cos(\delta)}}$
with the likely field stars distribution. However, the mean radial velocity of the 
system is equal to -51.81 km/s, which is $not$ consistent with the recession 
velocity of the cluster ($\overline{RV_{cl}}\pm\overline{\sigma_{cl}}=-29.11\pm0.85$ km/s), 
and its location in the sky is rather distant from the cluster's center (16.2 arcmin).

Given the present observations the most likely interpretation is that star 171895 is a 
very metal-rich field star falling by coincidence just at the cluster's turn-off. 
The presence of an additional close stellar companion in the system 
(e.g. a 0.8 $M_{\odot}$ star on a $\sim100$ days orbit, and presumably a white dwarf since we did not detect
double peaks in the cross-correlation function) would be required to reconcile the disagreement 
between the mean radial velocity of this star and the recession velocity of the cluster. 
Such scenario appears quite unlikely, otherwise since additional radial velocity measurements are 
necessary to constrain the mass of the transiting objects, 
this further hypothesis can be automatically checked. The HARPS instrument will perfectly accomplish these 
tasks, since we expect a precision of 10 m/s with 1 h of integration time for this star.

\subsection{Star $5622_3$}

\begin{figure*}
\center
\includegraphics[width=15cm]{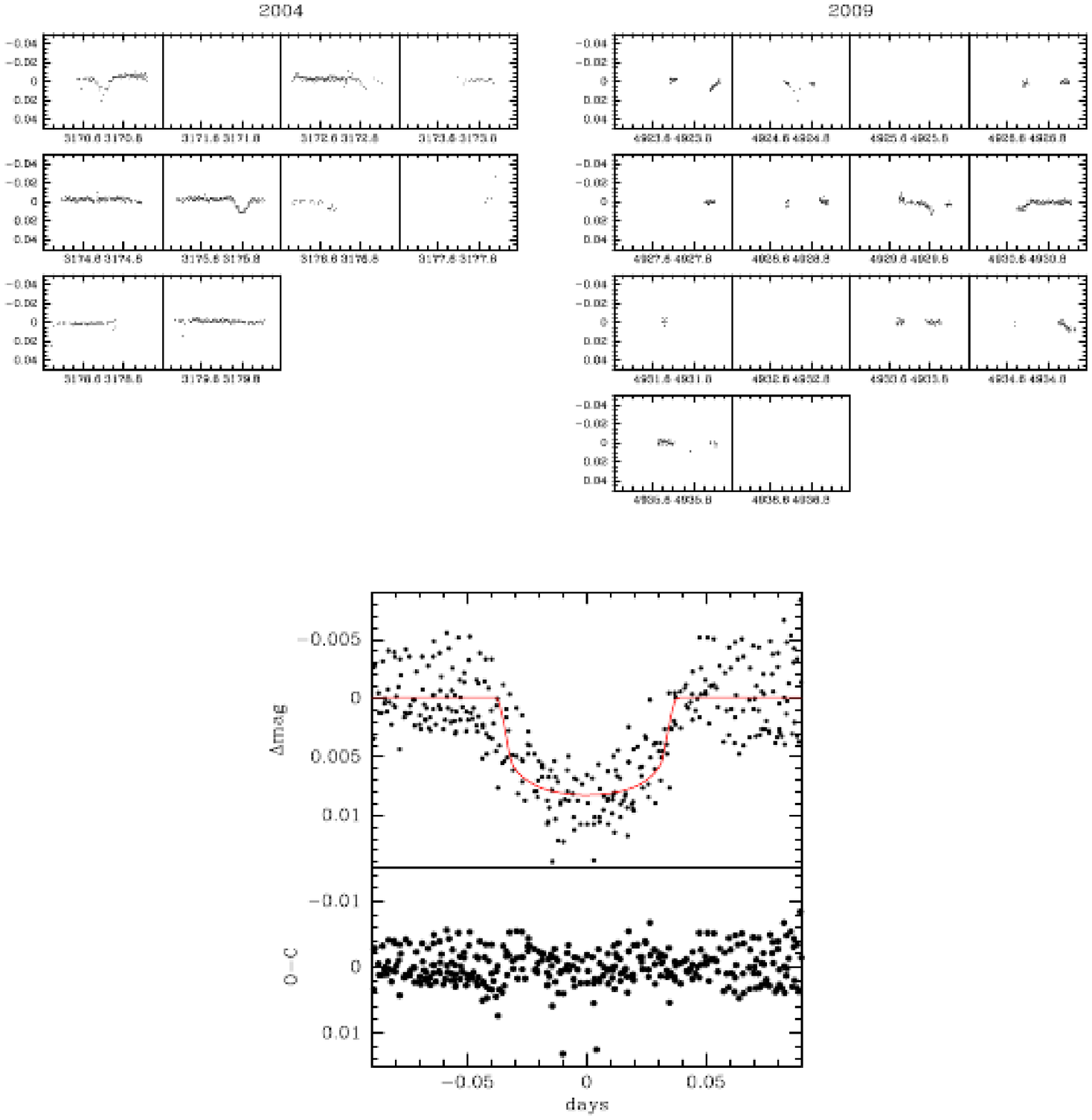}
    \caption{{\it Upper left:} lightcurve of star $5622_3$ relative to the
    2004 observing season. {\it Upper right:} lightcurve of star $5622_3$ relative
    to the 2009 observing season. {\it Lower panel, upper figure:} folded lightcurve. The continuous line
    indicates the best-fit model obtained using the Mandel \& Algol~(2002) algorithm,
    assuming that the host star is a solar type star. {\it Lower panel, bottom figure:} observed 
    minus model residuals. 
        }
\label{fig:cand_5622}
\end{figure*}

The magnitude of the host star is $V=17.789$. Its radial distance from the cluster's center
is 18.5 arcmin. In our photometry we detected 8 transit events
with a periodicity of $0.8494083$ days, (Fig.~\ref{fig:cand_5622}, upper panels). No UVES 
measurements were acquired for this system. This was probably due to a bad positioning
of the fiber, considering also the faintness of the object. The transits depth is 
$\sim0.008$ mag. Assuming that the host is a solar-type star we obtained the best-fit 
shown by the red line in Fig.~\ref{fig:cand_5622} (bottom panel). The RMS of the fit is 0.003 mag,
the radius of the transiting object is $R=$0.82 $R_{jup}$, and the inclination is 85$^{\circ}$.
We observe that the ingress and egress phases do not appear to 
closely follow the model, although the large scatter does not allow to draw a definitive conclusion.
The epoch ($E$) of the eclipses is given by:

\begin{displaymath}
(HJD - 450000.)\,=\,3168.99730\,+\,0.8494083\,\times\,E.
\end{displaymath}

\noindent
Since we are not able to characterize spectroscopically the host star, no additional 
informations on this system are provided here.

\subsection{Star 23333}

\begin{figure*}
\center
\includegraphics[width=15cm]{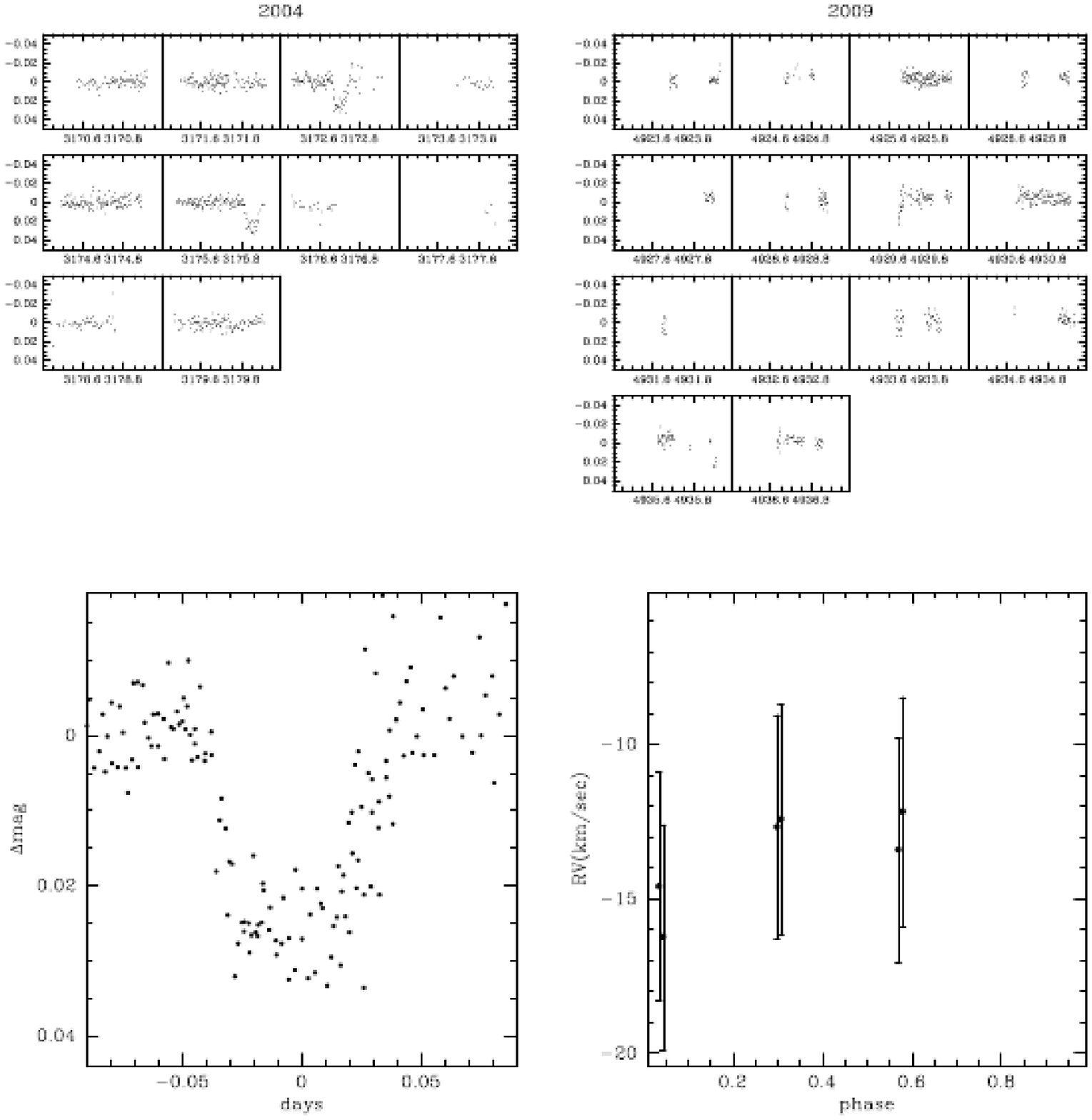}
    \caption{{\it Upper left:} lightcurve of star $23333$ relative to the
    2004 observing season. {\it Upper right:} lightcurve of star $23333$ relative
    to the 2009 observing season. {\it Lower left:} folded lightcurve. 
    {\it Lower right:} UVES spectroscopic measurements.
        }
\label{fig:cand_23333}
\end{figure*}

The magnitude of the host star is $V=18.247$. This object is the closest to the 
cluster's center, with a radial distance of 6.4 arcmin. From our photometry
we detected 4 transits events, 2 full transits in 2004 and 2 partial transits in 2009
(Fig.~\ref{fig:cand_23333} upper and lower left panels).
Given the faintness of the object the lightcurve is very noisy.
The transits are rather deep, being around 0.03 mag, and the duration is $\sim1.92$ hours.
The folded radial velocity measurements 
are shown in the bottom right panel of Fig.~\ref{fig:cand_23333}.
In this case radial velocity measurements would be sufficiently well separated in 
phase space to derive some orbital constraints. However, the
error of each measurement is large ($\sim3.6$ km/s). Even when binning the
couple of datapoints we acquired at the same orbital phase, the error is 
reduced by a $\sqrt\,2$ ($2.5$ km/s). More importantly, from the
UVES spectroscopy we were able to derive an approximate value of the effective
temperature of the host star, which is $T_{eff}=(5700\pm200)$ K. We were not able 
to derive the gravity since the low S/N of the spectra. However, given that effective
temperature, and assuming that the host is a dwarf star, we obtained that the
transits are too deep to be determined by a planetary body. The best fit obtained
with the Mandel \& Algol (2002) algorithm, considering a solar type host star, would
give a radius for the secondary equal to 2.25 $R_{jup}$ and an inclination of $84^{\circ}$. 
This is sufficient to exclude this object from the list of planetary 
transiting candidates.

\begin{table}
\caption{
Journal of UVES data for star 23333.
\label{tab:UVES_EB}
}

\begin{center}
\begin{tabular}{c c c}
\hline
\hline
Epoch & HRV(km/sec) & $err_{HRV}$(km/sec) \\
\hline
2454695.56104 & -16.254 &   3.627 \\
2454695.53224 & -14.608 &   3.710 \\
2454699.50991 & -12.687 &   3.627 \\
2454699.53824 & -12.442 &   3.710 \\
2454703.54373 & -12.193 &   3.710 \\
2454703.51466 & -13.425 &   3.627 \\
\hline
\end{tabular}
\end{center}
\end{table}

Among the three planetary transiting candidates presented, star 171895 is the 
most interesting object, both for its brightness and for its characteristics. Follow-up 
observations of this system (both photometric and spectroscopic in particular with 
the HARPS instrument) are clearly warranted to accurately determine its properties.

\section{Detached eclipsing binary systems}
\label{s:EA}

Among the objects in our target list there are four known detached 
eclipsing binary systems (see Sect.~\ref{s:introduction}). These objects were selected
from the list compiled by De Marchi et al.~(2010). As it appears evident from their Table~A.3
several of these detached systems present very shallow eclipses. In particular, six
objects in their list have eclipses with aplitudes $\le0.03$ mag. We decided to target with
the GIRAFFE spectrograph star 31195, 126376, 38138, and 24487, which have
eclipses of amplitudes 0.02 mag, 0.02 mag, 0.01 mag, and 0.53 mag, and periods
equal to 1.8156 days, 1.5896 days, 0.66606 days, and 0.85120 days respectively. The lightcurves
of these objects were already presented in De Marchi et al.~(2010). They tipically
show out of eclipse modulations, with the presence of primary and secondary eclipses.
The eclipses are also markedly V-shaped. The only exception may be star 38138, 
where the eclipses are shallow and noisy and not
easily distingishable from each others. Assuming that the obseved transits are grazing eclipses caused by very
close stellar companions (given the short periods), we should expect to observe large radial
velocity variations. For example, a stellar companion just at the limit between the 
brown dwarf and the stellar regime (M=0.08\,M$_{\odot}$) in a circular orbit with period equal to
$P=1.8156$ days would produce a radial velocity semi-amplitude equal to $K=12.5$ km/s
(assuming the primary star is a solar type star). From Table~\ref{tab:rv_fld} we see instead that
only star 24487 has been detected as a binary in our spectroscopic survey. In other 
words all the three objects presenting shallow eclipses did not appear radial
velocity variables. For star 38138 the observations have been acquired almost
at the same phase (0.16, 0.17, 0.19 in chronological order and assuming the time of
minimum and the period reported in Table~A.3 of De Marchi et al.~2010). For star 127376 the phase is also close to 0 or 0.5
(0.06, 0.58, 0.1). However, it appears that for star 31195 we should have reasonably expected
to detect a large radial velocity variations since the observations were acquired at orbital
phases equal to 0.11, 0.31, and 0.52. One possible explanation is that star 31195 is in fact
a hierarchical triple system, or a blend, where a third unresolved companion
determined shallower eclipses, and fictitiously small radial velocity variations. Since,
as we discussed in Sect.~\ref{s:results} and also pointed out by De Marchi et al.~(2010), 
this object might be considered a likely cluster's member despite its proper motion
membership probability is not very high, it would be very important
to acquire other observations to better clarify its nature.

\section{The eclipsing binary system 45368}
\label{s:EB}

\begin{figure}
\center
\includegraphics[width=8cm]{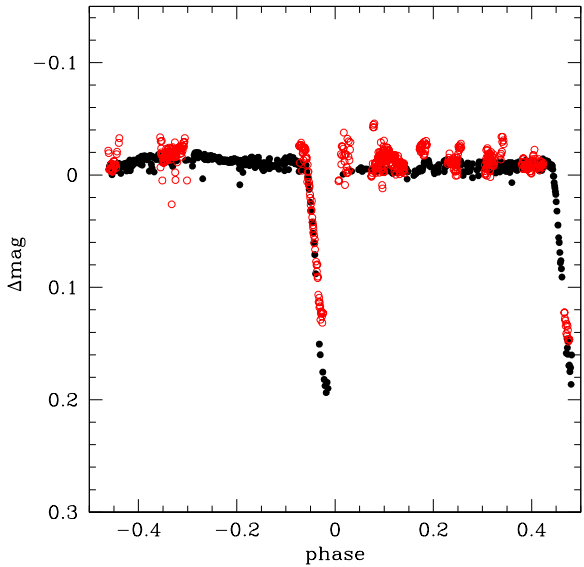}
    \caption{Folded light curve of the eclipsing binary system 45368. Black
             filled points are relative to the 2004 observing season, and red open
             circles to the 2009 observing season. 
        }
\label{fig:lctot_EB}
\end{figure}

Star 45368 is a double-lined eclipsing binary system located at the sub-giant branch 
of \object{NGC 6253}, and it is a very likely cluster's member as indicated by its proper
motion and position in the CMD (see Sect.~\ref{s:results}). In Fig.~\ref{fig:lctot_EB}, 
we present the photometric observations acquired for this object,
distinguishing between 2004 (black points) and 2009 (open red circles) observations. 
We detected in total four eclipses, all of them partial during the ingress of the 
transits. Moreover, as seen in Fig.~\ref{fig:lctot_EB}, the lightcurve is not constant out of 
the eclipses, since there are some clear light modulations. The 2009 observations
appear also in this case of lower quality with respect to the observations
acquired in 2004.

\begin{figure}
\center
\includegraphics[width=8cm]{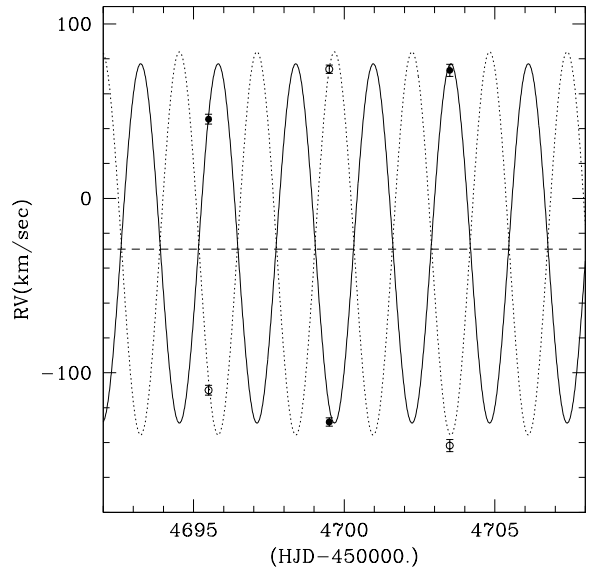}
    \caption{Fill black circles and open circles are UVES radial velocity measurements of the primary and 
             secondary components of the 45368 binary system. The solid curve is the best fitting
             model in phase with the primary eclipses. The dotted line is the best fitting model
             in phase with the secondary eclipse. The dashed horizontal line indicates the 
             baricentric velocity of the system ($\gamma=-30.19\,$km/s).
        }
\label{fig:EB_spec_fit}
\end{figure}

The UVES measurements presented in Fig.~\ref{fig:EB_spec_fit} indicate a very large
radial velocity variation ($\sigma_{obs}/\overline{\sigma_{obs}}=363$) consistent with
the idea of a stellar binary system and a short orbital period. Filled circles represent
the heliocentric radial velocities of the primary star, and open circles of
the secondary. The data are presented in Table~\ref{tab:UVES_EB}.
Our result for the epoch (E) of the primary eclipse is:

\begin{displaymath}
(HJD - 450000.)\,=\,3178.2903(1)\,+\,2.57317(1)\,\times\,E
\end{displaymath}

\begin{table*}
\caption{
Journal of UVES data for the 45368 eclipsing binary system.
\label{tab:UVES_EB}
}

\begin{center}
\begin{tabular}{c c c c}
\hline
\hline
Epoch & HRV1(km/s) & HRV2(km/s) & $\sigma_{HRV}$(km/s) \\
\hline
   2454695.53216 &   45.43 & -109.96 & 2.809\\
   2454699.53815 & -128.21 & 74.08 & 2.417\\
   2454703.54365 &   73.40 & -141.71 & 3.464\\
\hline
\end{tabular}
\end{center}
\end{table*}

\noindent
In Fig.~\ref{fig:EB_spec_fit}, we fit the radial velocity measurements with the following models:

\begin{displaymath}
RV1_{sim}(km/s)=0.2\,\,\,m2\,sin(i)\,\,P^{-1/3}(m1+m2)^{-2/3}\,\times
\end{displaymath}

\begin{equation}
\,\,\,\,\,\,\,\,\,\,\,\,\,\,\,\,\,\,\,\,\,\,\,\,\,\,\times\,cos\Big(\frac{2\pi}{P}\,(t-3178.2903470538)+\frac{\pi}{2}\Big)\,+\,\gamma
\end{equation}

\begin{displaymath}
RV2_{sim}(km/s)=0.2\,\,\,m1\,sin(i)\,\,P^{-1/3}(m1+m2)^{-2/3}\,\times
\end{displaymath}

\begin{equation}
\,\,\,\,\,\,\,\,\,\,\,\,\,\,\,\,\,\,\,\,\,\,\,\,\,\,\times\,cos\Big(\frac{2\pi}{P}\,(t-3178.2903470538)+\frac{3}{2}\pi\,\Big)\,+\,\gamma
\end{equation}

\noindent
where $m1$ and $m2$ are the masses of the primary and the secondary stars both
in solar masses, $P$ is the period in days, $t$ is the epoch of the UVES observations in days, 
$\gamma$ is the barycentric velocity of the system, $i$ the inclination of 
the system (assumed equal to 90$^{\circ}$), and the result is given in km/s. 
We assumed a circular orbit since the short orbital period should imply tidal circularization. 
The two radial velocity curves represented by Eq.~6 and Eq.~7 are in phase 
with the observed times of the primary and secondary eclipses respectively. We varied the two masses 
between 0.6$\,M_{\odot}$ and 10.5$\,M_{\odot}$ in steps of $0.1\,M_{\odot}$, and $\gamma$ between
$\pm\,4$ km/s from the recession velocity of the cluster ($-29.11$ km/s) in steps of 40 m/s.
The best solution was found minimizing the quantity:

\begin{displaymath}
\chi^2=\frac{\sum_{i=1}^{i=3}(RV1_i-RV1_{sim,i})^2/\sigma^2_{RV_i}}{N-3}+
\end{displaymath}

\begin{equation}
\,\,\,\,\,\,\,\,\,\,\,\,\,\,\,\,\,\,\,\,\,\,\,\,\,\,+\frac{\sum_{i=1}^{i=3}(RV2_i-RV2_{sim,i})^2/\sigma^2_{RV_i}}{N-3}
\end{equation}

\noindent
where $RV1_{i}$, $RV2_{i}$, are the observed radial velocities of the primary and of the
secondary respectively, $\sigma_{RV_i}$ are the uncertainties of the radial velocity measurements
as reported in Table~\ref{tab:UVES_EB} and $N-3=3$ since we have six independent measurements, and
three degrees of freedom ($m1$, $m2$, and $\gamma$). The result of the fit is represented by the 
solid and the dotted curves in Fig.~\ref{fig:EB_spec_fit} relative to the primary and
secondary star respectively, and implies $m1\sim1.6\,M_{\odot}$, $m2\sim1.5\,M_{\odot}$,
$\gamma=-30.2$ km/s, and $\chi=1.23$. The value of the barycentric velocity of the system is then
fully consistend with cluster's membership, leaving little doubt that the system belongs to
the cluster. The masses we derived confirm also the idea that these stars are cluster's evolved objects
as suggested by the position along the sub-giant branch of the cluster, and by the fact
that turn-off stars have masses around $\sim1.3\,M_{\odot}$, as derived by isochrone fitting.

In Table~\ref{tab:results_EB}, we summarize the results obtained from our analysis. Since 
in our photometry the two eclipses are only partial, we did not try to fit the lightcurve of 
this system and constrain the radii of the stars. We note that between the 2004 
and 2009 measurements there could be a photometric offset.
However, a complete and detailed analysis of this system will be presented once more data will be available.
The possibility to derive accurate masses and radii for these evolved stars will offer an
excellent oppotunity to test stellar evolution models predictions at the extreme metallicity
of \object{NGC~6253}. 

\begin{table}
\caption{
Summary of the parameters derived for the 45368 eclipsing binary system,
\label{tab:results_EB} assuming $i=90^{\circ}$ and $e=0$. 
}

\begin{center}
\begin{tabular}{c c}
\hline
\hline
Period(days) & 2.57317(1) \\
HJD of primary eclipse & 2453178.2903(1) \\
$\gamma$(km/s) & -30.19 \\
$m2/m1$ & 0.94 \\
$m1(M_{\odot})$ & $\sim$1.6 \\
$m2(M_{\odot})$ & $\sim$1.5 \\
\hline
\end{tabular}
\end{center}
\end{table}

\section{Radial velocity searches for sub-stellar companions around old open clusters turn-off stars}
\label{s:RV_search}
 
In this Section, we estimated the number of planets we can expect to detect 
during a radial velocity survey toward \object{NGC~6253}, considering
different observing strategies. We focused our attention on cluster's turn-off stars ($V=15$), and assumed to
use the HARPS spectrograph. At the magnitude of our target stars, considering
1 h of integration time for each star, we expect a radial velocity precision of 10 m/s per measurements. 
 A 3-$\sigma$ detection threshold implies that only substellar companions inducing radial velocity 
semi-amplitudes variations $K>30$ m/s could be detected. From the 
Fischer \& Valenti~(2005) law we expect $\sim18\%$ $FGK$ dwarf stars having planets producing radial velocity semi-amplitudes $K>30$ m/s,
and with orbital periods up to 4 years, at the metallicity of \object{NGC~6253} ([Fe/H=+0.39). We used the
following Equation to simulate the observations of a planet-host star at some fixed epochs: 

\begin{equation}
RV_{sim}(m/s)=200\,\,\,m\,\,\,sin(i)P^{-1/3}M^{-2/3}cos(n\tau+\phi)
\end{equation}

\noindent
where $n$ is the mean orbital motion ($=2\pi/P$), and $\tau$ corresponds to the 3 epochs 
of our observations. In the above Equation, the period is expressed in days, 
the mass of the companion ($m$) in Jupiter masses, the mass of the primary ($M$) in 
solar masses, and the resulting radial velocities are in meters per second. Cluster's stars with 
the same magnitude of our targets have masses of $\sim$1.3$M_{\odot}$ (derived from
isochrone fitting), and we adopted this value for the primary's mass. 
Random gaussian scatter was also added
to the measurements to account for observational errors (the dispersion was fixed to 10 m/s). The orbital period
of the planet was varied between 1 day and 1460 days ($=4$yr) in steps of 0.5 days, and the mass of the secondary
between 0.5 $M_{jup}$ and 13 $M_{jup}$, in steps of 0.1 $M_{jup}$. These limits are consistent with the validity
domain of the Fischer \& Valenti (2005) law. For each couple of period and mass we performed 1000 simulations randomly varing 
the orbital phase of the planet, and defined the detection efficiency ($eff$) as the ratio between the number of 
simulations for which the standard deviation of the simulated measurements around their mean value 
exceeded $3\times\sigma=30$m/s, with respect to the total number of simulations. 

\begin{figure*}
\center
\includegraphics[width=11.5cm]{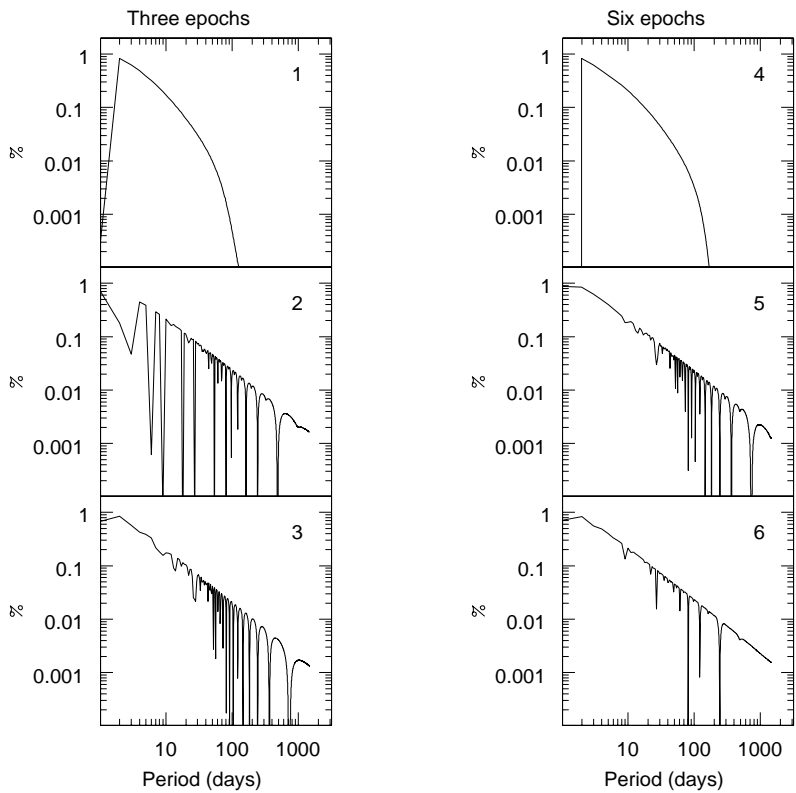}
    \caption{
             Probability to detect a planet around
             a single turn-off star of \object{NGC~6253}, in function of the planetary 
             orbital period and different simulated HARPS observing strategies.
             The left panels show the results
             considering three observing epochs and the right panels considering 
             six observing epochs, as reported by the 
             labels at the top of each group of 
             panels. Each small panel refers to one of the simulations presented in 
             Table~\ref{tab:results_simulations}, as indicated by the inner numeration. 
        }
\label{fig:results_simulations_period}
\end{figure*}

\begin{figure*}[!]
\center
\includegraphics[width=11.5cm]{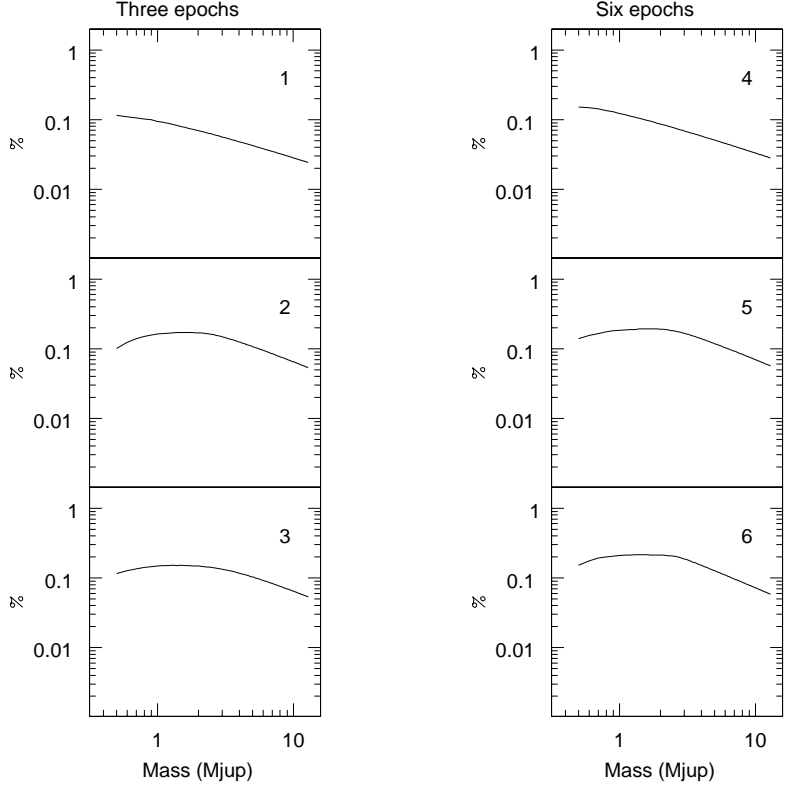}
    \caption{Same as Fig.~\ref{fig:results_simulations_period}, but in function of the planetary mass.
        }
\label{fig:results_simulations_mass}
\end{figure*}

In order to calculate the expected number of planets, we convolved the detection efficiency with the mass-period functions
of extrasolar planets as derived by Jiang, I.-G. et al.~(2010). We considered the results obtained by those authors 
for independent mass functions, and in particular for the case of the single imaginary survey. Assuming
coupled mass-period functions gives similar results, at least for the case of the single imaginary survey, as 
can be deduced from their Table~5. In such a way the probability ($dP$) that a single star of 
metallicity [Fe/H] has a planet with orbital period comprised between $P$ and $P+dP$, and mass comprised 
between $M$ and $M+dM$ is given by:

\begin{equation}
dP=c\,\left(\frac{M}{M0}\right)^{-\alpha}\,\left(\frac{P}{P0}\right)^{-\beta}\,\frac{dM}{M}\,\frac{dP}{P}\,\times\,10^{2[Fe/H]}\times\,eff(M,P)
\end{equation}

\noindent 
where $\alpha=-0.143$, $\beta=-0.124$, $c=0.001316$, $M0=1.5\,M_{jup}$, and $P0=90$ days. The efficiency function ($eff[M,P])$
depends not only on the mass and period of the orbiting planet, but also on the observing strategy.
In the above Equation we have assumed
that the mass-period functions are independent from the host-star's metallicity, and from the host-star's mass. 
Ribas \& Miralda-Escud\'e (2007) derived that the host star metallicities decrease with planet mass,
however the significance of this result is only marginal considering the present sample of exoplanets.
A dependence of planet's mass and orbital period on the planet's host mass has been also demontrated. In particular
around M-dwarf stars hot-jupiter planets appear relatively rare (e.g. Endl et al.~2006), 
whereas they seem more frequent around giant stars with respect to dwarf stars, having long orbital periods, 
and rather large minimum masses (e.g. Sato, B. et al. 2007). Our considerations are however limited 
to neptune-jupiter planets around dwarf stars at the cluster's turn-off, then are not affected by this problem.
 Once applied to the period-mass domain of our simulations (see above), assuming solar 
metallicity and perfect efficiency ($eff=1$), Equation 8 
gives a detection probability of $3.1\%$, in good agreement with the normalization factor of the Fischer \& Valenti (2005) 
law ($3\%$). Once particularized to our case, assuming the metallicity of \object{NGC~6253},
the above Equation allows to calculate the total number of detectable planets
and their distribution in period and mass, as  reported in Table~\ref{tab:results_simulations} and 
shown in Fig.~\ref{fig:results_simulations_period} and  Fig.~\ref{fig:results_simulations_mass}.
The main result is that the observing strategy has a large impact on the expected number of planets.
With 3 epochs equally distributed over 4 years, and assuming an initial sample of 26 stars
(which means a total of 78 HARPS hours) we can expect to detect 3.23 planets.
If taken consecutively, we could expect to detect only 1.42 planets.
In general, strategies that involve non consecutive observations are the most effective since they allow higher
sensitivity to longer period planets.

\begin{table}
\caption{
Number of detectable planets from different HARPS simulated observing runs.
The results below are given assuming a total number of surveyed 
turn-off stars of \object{NGC~6253} equal to 26.
\label{tab:results_simulations}
}
\begin{center}
\begin{tabular}{c c c}
\hline
\hline
N.sim & Epochs (days) & N.Planets \\
\hline
1 & t1=1;t2=2;t3=3 & 1.42 \\
2 & t1=486;t2=972;t3=1458 & 3.23 \\
3 & t1=1;t2=2;t3=730 & 3.06 \\
4 & t1=1;t2=2;t3=3;t4=4;t5=5;t6=6 & 1.73 \\
5 & t1=1;t2=2;t3=729;t4=730;t5=1459;t6=1460 & 3.58 \\
6 & t1=1;t2=243;t3=486;t4=729;t5=972;t6=1215 & 3.87 \\
\hline
\end{tabular}
\end{center}
\end{table}

\section{Minimum mass and period of companion objects of cluster's members close binary systems}
\label{s:simulation}

In this Section we studied the opposite problem of Section~\ref{s:RV_search}:
assuming to find a radial velocity variable star during an $RV$ survey and
assuming that this variation is determined by the presence of a companion object,
which are the most likely minimum mass ($m\,sin[i]$) and period ($P$) of this object?
The effectiveness of an observing strategy should be evaluated not only considering if
it is able to maximize the number of detectable planets, but also considering if it 
is able to minimize the number of spurious objects that may produce the same signal of the planets.

\begin{figure}
\center
\includegraphics[width=8cm]{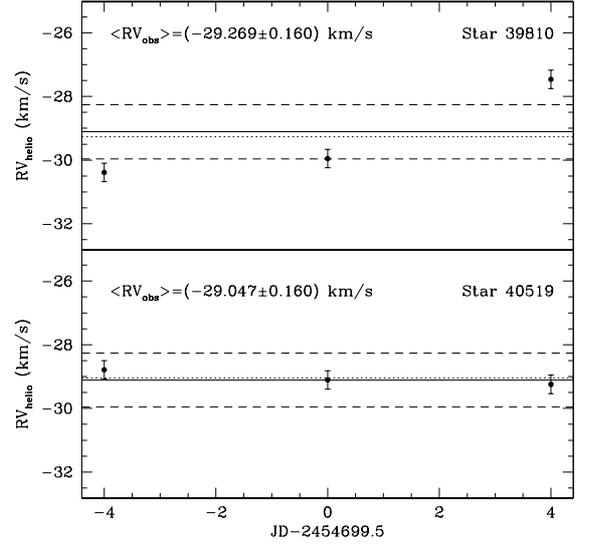}
    \caption{Radial velocities obtained for stars 39810 and 40519 located in the turn-off
             region. 
             Dotted lines indicate the stars' mean radial velocities
             ($\overline{RV_{obs}}$), solid lines represent the mean 
             radial velocity of the cluster ($\overline{RV_{cl}}$), and the dashed lines 
             represent the 1 $\overline{\sigma_{cl}}$  
             uncertainty range of the cluster's mean radial velocity ($\overline{\sigma_{cl}}=0.85$ km/s, Sect.~\ref{s:results}).
        }
\label{fig:cand}
\end{figure}

As an illustrative example, we considered at first the case of the $UVES$ observations, and in particular of 
star 39810, which has the smallest dispersion among our sample of cluster's candidate close binary systems,
as shown in Sect.~\ref{s:results}. In Fig.~\ref{fig:cand}, we present a comparison between 
the radial velocity measurements of star 39810 and of star 40519, which is also a turn-off cluster's member 
and not a radial velocity variable star. In this Figure,
dotted lines indicate the stars' mean radial velocities ($\overline{RV_{obs}}$), solid lines represent the mean 
radial velocity of the cluster ($\overline{RV_{cl}}$), and dashed lines 
represent the 1 $\overline{\sigma_{cl}}$ uncertainty range of the cluster's mean radial velocity 
($\overline{\sigma_{cl}}=0.85$ km/s, Sect.~\ref{s:results}). While star 40519 radial velocities
are all consistent with the cluster's recession velocity, measurements for star 39810
clearly depart from it. We assume in the following that the observed radial velocity variation for star 39810
is determined by the presence of a companion object. On the basis of this hypothesis, we constrain the expected
minimum mass and period of this object by means of Monte Carlo simulations. The basic condition
that we imposed is that the barycentric radial velocity of the binary system is coincident with
the recession velocity of the cluster. This is possible thanks to the fact that
this star is a very likely cluster's member, as demonstrated in Section~\ref{s:results}. 
Was this object a common field star, we wouldn't have had any a priori knowledge on its barycentric radial velocity. 

\begin{figure}
\center
\includegraphics[width=8cm]{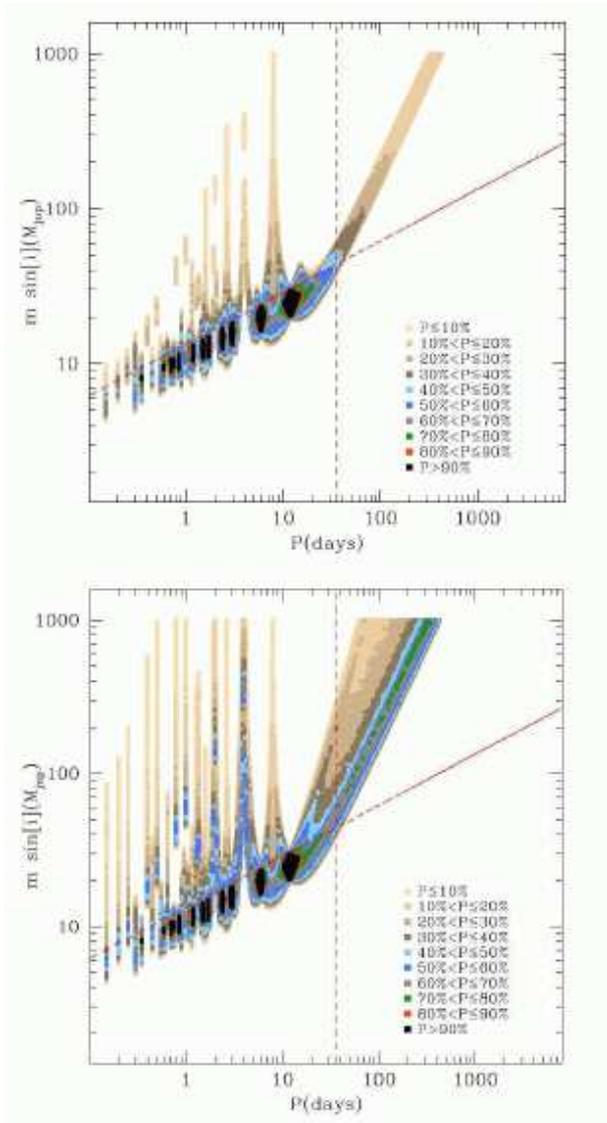}
    \caption{Period-mass probability distribution for the possible companion object of star 39810, 
             obtained from orbital simulations, assuming the real observing epochs and circular obits. 
             In the upper panel we show the results in the hypothesis that the star belongs to the
             cluster (as demonstrated by proper motions, radial velocities and position in the CMD), 
             whereas in the bottom panel we show the result if the star was a common field
             object. $\sigma_{obs}=1595$ m/s,  $\overline{\sigma_{obs}}=304$ m/s,
             $\overline{\sigma_{cl}}=850$ m/s.
        }
\label{fig:prob}
\end{figure}

We varied the period $P$ of the orbit between 0.1 to 1000 days 
($dP=$0.05 days), and the minimum mass of the companion ($m\,sin[i]$) between 0.1 $M_{\odot}$ to 1000 $M_{jup}$ 
($dm\,sin[i]$= 0.05 M$_{jup}$). We assumed circular orbits, and for each 
fixed value of the mass and the period we performed $100$ simulations 
randomly chosing the orbital
phase between $0<\phi<2\pi$. Then we calculated the observed radial 
velocity of the primary star ($RV_{sim}$) by means of Equation~9
and, for each simulation, we calculated the 
dispersion of the radial velocity measurements $\sigma_{obs,sim}$, and the mean radial 
velocity $\overline{RV_{sim}}$. We finally imposed the conditions that the simulated dispersions
and mean radial velocities matched our observed values:

\begin{equation}
\sigma_{obs}-\overline{\sigma_{obs}}\le\sigma_{obs,sim}\le\sigma_{obs}+\overline{\sigma_{obs}},
\end{equation}

\begin{equation}
|\overline{RV_{obs}}-\overline{RV_{cl}}|-\overline{\sigma_{cl}}\le|\overline{RV_{sim}}-\overline{RV_{cl}}|\le|\overline{RV_{obs}}-\overline{RV_{cl}}|+\overline{\sigma_{cl}},
\end{equation}

\noindent
where $\overline{\sigma_{obs}}$ is our adopted radial velocity error ($\overline{\sigma_{obs}}=0.304$ m/s, see Eq.~3). 
The likelihood a given period and minimum companion mass imply orbits for which the simulated mean radial 
velocities and dispersions are consistent with our observations, was calculated dividing the 
number of simulations that matched our criteria, to the total number 
of simulations. In Fig.~\ref{fig:prob} (upper panel), we show the likelihood 
distribution we obtained for star 39810. 
For periods $P<3$ days, the mass is m$\,$sin(i)$\lesssim$13 $M_{jup}$, then in the range of massive jupiters.
We also plot the theoretical curve (dashed red line) correspondent to $K=1.4\times\sigma_{obs}$, which links 
the observed scatter to the expected radial velocity amplitude of the companion, and the vertical dashed red line
indicating a period equal to $\sim4\times[t3-t1]=36$ days, which appears to be the maximum period we should
expect for the companion object (coincident with a maximum mass 
equal to m$\,$sin(i)=40$M_{jup}$), at a confidence level $>50\%$.
In Fig.~\ref{fig:prob} (bottom panel), we show the probability distribution we obtained
for the same object, but without the assumption on the barycentric radial velocity
of the system (as if the star was a field object). Regions with probability larger than $50\%$
clearly extend toward larger periods and minimum masses ($P>36$ days, m$\,$sin(i)$\,>40\,M_{\odot}$).

\begin{figure}
\center
\includegraphics[width=8cm]{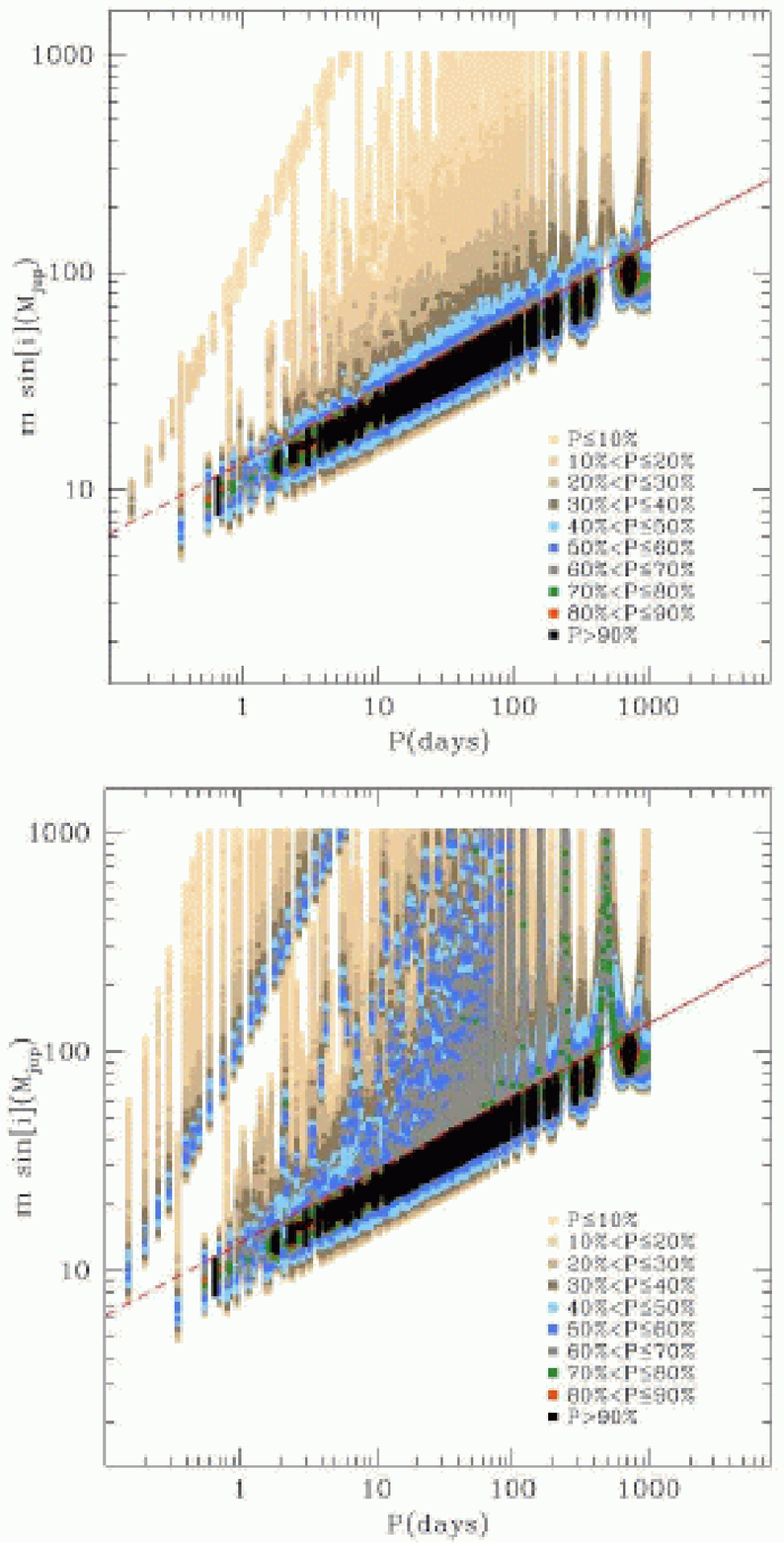}
    \caption{
            Same as Fig.~\ref{fig:prob}, assuming three observing epochs at t1=486 days,
            t2=972 days, and t3=1458 days and circular orbits, $\sigma_{obs}=1595$ m/s,
            $\overline{\sigma_{obs}}=304$ m/s, $\overline{\sigma_{cl}}=850$ m/s.
        }
\label{fig:prob1}
\end{figure}

To better understand this result we performed other simulations, assuming different observing epochs.
In Fig.~\ref{fig:prob1}, we show the results considering three epochs at $t1=486$ days, $t2=972$ days, 
and $t3=1458$ days (equally distributed over a period of four years). 
In both cases (cluster and field star) the overall shape of the probability distribution is 
the same as in Fig.~\ref{fig:prob}, however the highest probability regions are clearly more extended toward
larger masses and periods. The interpretation of  Fig.~\ref{fig:prob} and
 Fig.~\ref{fig:prob1} is that as long as the period of the companion object is probed by the timescale
of the observations (where the timescale can be roughly assumed equal to $4\times[t3-t1]$), the most likely mass and
period of the companion object producing the observed radial velocity scatter are found along the theoretical
curve both for the case of the cluster and of the field star (as traced by the highest probability regions). However,
a worse result is obtained for the case of the field star in correspondence of resonant periods. Objects
with periods longer than the observing timescale may as well produce a scatter compatible with the observations,
but their mass should be larger than what implied by the theoretical curve, since they are forced to produce 
that scatter during the timescale spanned by the observations. This determines the tail toward large masses and periods
visible in Fig.~\ref{fig:prob}. Otherwise, considering the case of the cluster's
star, this tail is reduced in extension with respect to the case of the field star, and implies probabilities $<30\%$ 
thanks to the condition on the barycentric velocity of the system (against $<80\%$ in the case of the
field star).

The major conclusion is that if a star belongs to a cluster, and our observations
span a period $\Delta\,T$ during which we determine a radial velocity scatter equal 
to $\sigma_{obs}>\sigma_{cl}$, and a mean radial velocity consistent with the recession velocity
of the cluster, the expected mass and period of the suspected companion object 
are comprised in the region defined by the conditions $K=1.4\times\sigma_{obs}$, and 
$P\lesssim4\times\,\Delta\,T$ in the mass-period diagram, at a confidence level $>50\%$.
This is not true for the case of the field star, where mass and period can be 
larger than these limits considering the same confidence level.

\subsection{The best observing strategy}

\begin{figure}
\center
\includegraphics[width=8cm]{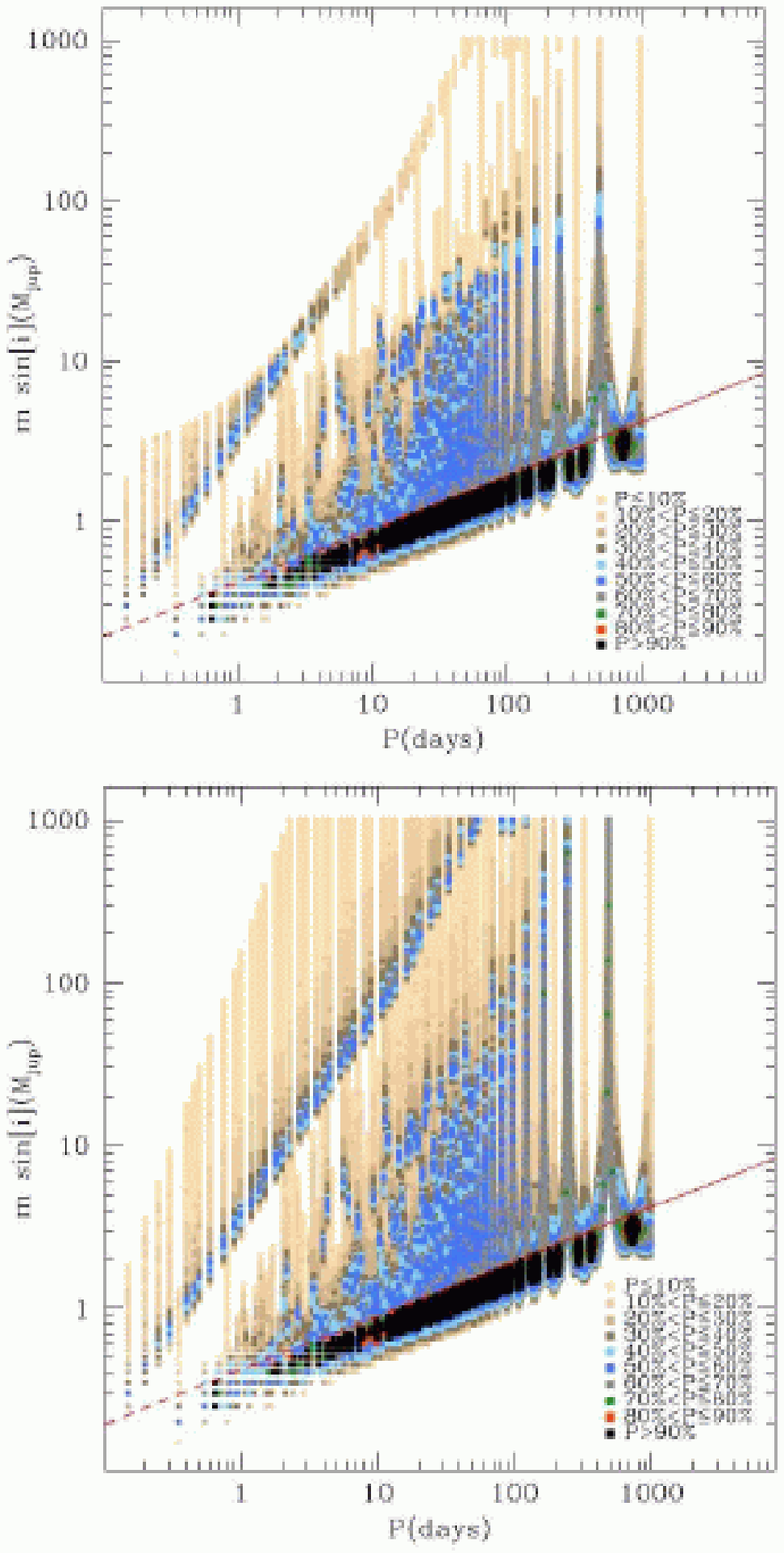}
    \caption{
            Same as Fig.~\ref{fig:prob}, assuming three observing epochs at t1=486 days,
            t2=972 days, and t3=1458 days, circular orbits, $\sigma_{obs}=50$ m/s,
            $\overline{\sigma_{obs}}=10$ m/s, $\overline{\sigma_{cl}}=850$ m/s.
        }
\label{fig:prob_harps}
\end{figure}

In Section~\ref{s:RV_search}, we derived that three observing epochs equally distributed over a period of four years 
should provide a good compromize between the number of detectable planets in \object{NGC~6253} and 
observing time (Table~\ref{tab:results_simulations}). In order to understand how efficient can be 
this observing strategy to isolate potential planetary candidates, we assumed to observe with the HARPS
instrument ($\overline{\sigma_{obs}}=10$ m/s), fixing the observing epochs at t1=486, t2=972 and
t3=1458 days. We assume to study a star presenting a radial velocity scatter $\sigma_{obs}=50$ m/s. 
Period and mass of the companion object where varied 
between 0.05 $M_{jup}$ and 1000 $M_{jup}$, and between 0.05 days and 1000 days 
respectively ($dP\,=\,0.05$ days, $dm\,sin(i)\,=\,0.05\,M_{jup}$). 
As shown in Fig.~\ref{fig:prob_harps}, the result for the cluster's star resemble more
the case of the field star than in previous simulations. The reason is that in this 
case $\sigma_{obs}<<\overline{\sigma_{cl}}$ which
reflects the intrinsic dispersion of radial velocities of cluster's stars, then the condition
on the mean radial velocity is not as efficient in constraining the period and mass
of the orbiting companion as for larger observed $\sigma_{obs}$ values.
Still it is clear that a better result can be obtained than in the case of a field object.

From our survey we have detected five stars having mean radial velocities 
consistent with the cluster's recession velocity within
0.755 km/s, as shown in Table~\ref{tab:top_targets}.
Moreover they are all cluster's turn-off stars, very likely cluster's members on the basis of our
proper motions and radial velocities, and do not present signs of photometric and 
spectroscopic variability. These objects are: star 40519, 45512, 44104, 45523, and 45267. 

Star 39810 discussed above may be instead more properly followed-up using the same 
UVES/GIRAFFE spectrographs.

We note that also star 45387 has a small radial velocity dispersion, and it is a radial
velocity variable, although this object is located in the 
blue straggler region, and the assumption on the mass of the primary may not be 
correct. The other close binary systems in our sample are more likley to host stellar companions.

\begin{table}
\caption{
Turn-off stars of \object{NGC~6253} which are optimal targets for sub-stellar
companion searches.
\label{tab:top_targets}
}

\begin{center}
\begin{tabular}{c c c}
\hline
\hline
ID$_{MO}$ & $\overline{RV_{obs}}-\overline{RV_{cl}}$ \\
         & (km/s) \\
\hline
40519 & 0.063 \\
45512 & 0.111 \\
44104 & 0.165 \\
45523 & 0.277 \\
45267 & 0.755 \\
\hline
\end{tabular}
\end{center}
\end{table}

Searching for sub-stellar objects (brown dwarfs and jupiter planets) 
with the radial velocity technique around old open clusters turn-off stars appears then 
feasible also with present day instrumentation, but it is necessary: (i) an efficient pre-selection 
of candidate cluster's members (by means of colors, magnitudes, and possibly proper motions, 
and radial velocities); (ii) further identification of massive sub-stellar objects, a task that
can be accomplished using multi-object spectrographs like UVES/FLAMES. With just a few hours of 
observation (6 hours in our case), it is possible to easily cover most of the 
turn-off members, and identify the best brown dwarf, massive planet candidates, providing at the 
same time a list of cluster's stars that do not appear radial velocity variables at the precision of these instruments; 
(iii) this sample, $cleaned$ of spurious field contaminants and of large mass-ratio binary systems,
can be followed-up with high precision spectroscopy, at first using the same technique of UVES/FLAMES,
and then covering more intensively only the most promising targets. The previous 
steps constitute a sort of roadmap towards detection
of sub-stellar objects in open clusters. Proceding by steps, allows to make a
more efficient use of observational time.

\section{Conclusions}
\label{s:conclusions}

We have presented the results of the first multi-epoch radial velocity survey toward
the old metal-rich open cluster \object{NGC 6253}. The mean radial velocity of the cluster
is $\overline{RV_{cl}}\pm\overline{\sigma_{cl}}=(-29.11\pm0.85)$ km/s. Using our photometry,
proper motions, and radial velocities we identified
35 likely cluster's members, populating the turn-off, sub-giant, red-giant, red clump, and blue
straggler regions of the cluster. Among this sample, 12 objects are likely cluster's close binary
systems. 

We detected one object that may have
a companion with a minimum mass in the sub-stellar regime, and that needs to be further 
investigated: star 39810.

We isolated five turn-off stars that are optimal targets to search for planetary mass companions 
with the high precision spectroscopy: stars 40519, 45512, 44104, 45523, 45267.

Among the three planetary candidtes we found in the region of \object{NGC~6253}, star 
171895 is the most interesting objects, requiring a more intensive photometric and spectroscopic follow-up
in particular with high precision spectroscopy.

We found a new sub-giant branch eclipsing binary system: star 45368.

We compared our results with previous literature radial
velocity measurements of a few cluster's members, finding in general a good agreement with our study. 

The close binary frequency among turn-off and evolved cluster's stars is
f$_{cl}=$(29$\pm$9)$\%$ (f$_{bin}=$[34$\pm$10]$\%$,  once including also
longer period binaries), appears larger than the field binary frequency
f$_{cl,field}=$(22$\pm$5)$\%$, although still consistent with that estimate
considering the uncertainties.

We showed that searching for sub-stellar objects (brown dwarfs and jupiter planets) 
with the radial velocity technique around old open clusters turn-off stars, appears feasible also with 
present day instrumentation, but it is necessary: (i) an efficient pre-selection 
of candidate cluster's members; (ii) further identification of likely massive sub-stellar objects
by means of multi-object spectroscopy; (iii) subsequent follow-up of the best 
target stars by means of high precision spectroscopy. 
We derived that three observing epochs uniformly distributed over a period of four years
give a good compromise between expected number of detectable planets, and observing time.

In Table 12 we report the cross-correlation among the star's ID introduced in Montalto et al.~(2009)
and other authours and catalogs.

Very recently Anthony-Twarog et al.~(2010) have presented new radial velocity data
on \object{NGC~6253} members. These measurements have not been included in our analysis. Merging
the results presented in that work with our own results, will allow 
further identification of the best targets to follow-up more intensively with high resolution
spectroscopy to search for sub-stellar companions around cluster's turn-off stars.

\acknowledgements{
The authors are grateful to the anonynous referee for the numerous useful
comments which have allowed a significant improvement of the paper.
}

{}

\begin{table}
\caption{
ID cross correlation table.
\label{tab:crosscorr}
}
\begin{center}
\begin{tabular}{c c c c}
\hline
\hline
ID$_{MO}$ & ID$_{DM}$ & ID$_{BR}$ & Spectr. Ref. \\
\hline
39621  & 21000$_2$ & 878          & \\
39866  & 21303$_2$ & 1704         & \\
39994  & 21462$_2$ & 2131 &\\
40010  & 21487$_2$ & 2238 &\\
44079  & 26902$_2$ &  904 &\\
44104  & 26943$_2$ & 1393 &\\
44714  & 28028$_2$ & 1027 &\\
45144  & 28976$_2$ & 4181 &\\
45267  & 29299$_2$ &  917 &\\
45285  & 29358$_2$ & 3699 &\\
45300  & 29398$_2$ & 7207 &\\
45368  & 29635$_2$ & 2242 &\\
45387  & 29695$_2$ & 1646 &\\
45403  & 29716$_2$ & 2696 &\\
45404  & 29718$_2$ & 3138 & Sestito et al.~(2007) \\
45412  & 29743$_2$ & 2509 & Carretta et al.~(2007) \\
45413  & 29744$_2$ & 2508 & Carretta et al.~(2007) \\
45421  & 29754$_2$ & 2542 & Sestito et al.~(2007) \\ 
45422  & 29755$_2$ & 2343 & \\
45453  & 29789$_2$ & 1830 & \\
45474  & 29811$_2$ & 2225 & Sestito et al.~(2007) \\
45495  & 29833$_2$ & 4227 & \\
45497  & 29835$_2$ & 2400 & \\
45512  & 29851$_2$ & 4391 & \\
45513  & 29852$_2$ & 2864 & \\
45523  & 29863$_2$ &  301 & \\
45524  & 29864$_2$ & 4308 & \\
45528  & 29868$_2$ & 2814 &\\
39810  & 21234$_2$ & 1562 &\\
40519  & 22119$_2$ & 4029 &\\
45414  & 29745$_2$ & 4510 & Carretta et al.~(2007) \\
45444  & 29779$_2$ & 2126 & \\
44682  & 27978$_2$ &  307 & \\
45447  & 29782$_2$ & 2726 & \\
45410  & 29279$_2$ & 2885 & Carretta et al.~(2007) \\
30341  &  9268$_2$ & 3272 &\\
  31195  & 10340$_2$ & 3955 &\\ 
 173273  & 11351$_8$ & - & \\
   9832  & 13461$_1$ & 7832 & \\
      -  & 15343$_2$ & 7502 & \\
  24487  &  1866$_2$ & 7006 & \\
  38138  & 19192$_2$ & - & \\
 134894  & 20096$_6$ & - & \\
  40819  & 22477$_2$ & 4987 & \\
 163536  & 23116$_7$ & - & \\
      -  & 23129$_7$ & - & \\
      -  & 23130$_7$ & - & \\
 163548  & 23133$_7$ & - & \\
      -  & 23134$_7$ & - & \\
 163549  & 23135$_7$ & - & \\
 163551  & 23139$_7$ & - & \\
 163552  & 23141$_7$ & - & \\
 163560  & 23150$_7$ & - & \\
 163574  & 23166$_7$ & - & \\
 163577  & 23169$_7$ & - & \\
 163588  & 23181$_7$ & - & \\
 163594  & 23187$_7$ & - & \\
\hline
\end{tabular}
\end{center}
\end{table}

\newpage

\addtocounter{table}{-1}          

\begin{table}
\caption{
- continued.
\label{tab:crosscorr}
}
\begin{center}
\begin{tabular}{c c c c}
\hline
\hline
ID$_{MO}$ & ID$_{DM}$ & ID$_{BR}$ & Spectr. Ref. \\
\hline
  16649  & 23368$_1$ & 7592 & \\
      -  & 25450$_2$ & 3332 & \\
  67886  & 26321$_3$ & - & \\
  67921  & 26378$_3$ & - & \\
  67922  & 26379$_3$ & - & \\
  67924  & 26381$_3$ & - & \\
      -  & 26383$_3$ & - & \\
      -  & 26384$_3$ & - & \\
  67926  & 26387$_3$ & - & \\
  67927  & 26388$_3$ & - & \\
  67928  & 26389$_3$ & - & \\
  67929  & 26391$_3$ & - & \\
  67930  & 26392$_3$ & - & \\
  67932  & 26394$_3$ & - & \\
  67934  & 26396$_3$ & 7062 & \\
  67936  & 26399$_3$   & - & \\
  67941  & 26404$_3$   & - & \\
  67945  & 26408$_3$   & - & \\
  67946  & 26409$_3$   & - & \\
  67947  & 26410$_3$   & - & \\
  67955  & 26418$_3$   & - & \\
  67966  & 26429$_3$   & - & \\
  67967  & 26430$_3$   & - & \\
  67977  & 26440$_3$   & - & \\
  67985  & 26448$_3$   & - & \\
  67990  & 26453$_3$   & - & \\
  68015  & 26480$_3$   & - & \\
 140621  & 28393$_6$   & - & \\
      -  & 28459$_6$   & - & \\
 140634  & 28460$_6$   & - & \\
 140638  & 28464$_6$   & - & \\
 140644  & 28473$_6$   & - & \\
 140646  & 28475$_6$   & - & \\
 140649  & 28478$_6$   & - & \\
 140657  & 28486$_6$   & - & \\
 140669  & 28498$_6$   & - & \\
 140673  & 28503$_6$   & - & \\
 140699  & 28529$_6$   & - & \\
      -  & 28690$_8$   & - & \\
      -  & 28694$_8$   & - & \\
 185675  & 28704$_8$   & - & \\
 185677  & 28706$_8$   & - & \\
 185680  & 28709$_8$   & - & \\
 185688  & 28717$_8$   & - & \\
 185691  & 28720$_8$   & - & \\
      -  & 28735$_8$   & - & \\
 185713  & 28743$_8$   & - & \\
 185716  & 28746$_8$   & - & \\
 185725  & 28755$_8$   & - & \\
  45392  & 29703$_2$   & 890 & \\
  45396  & 29707$_2$   & - & \\
     45409  & 29738$_2$   &  5201 & \\
     45427  & 29761$_2$   & - & \\
     45433  & 29768$_2$   & 7049 & \\
     45436  & 29771$_2$   & 1222 & \\
         -  & 29788$_2$   & 1444 & \\
\hline         
\end{tabular}
\end{center}
\end{table}

\newpage

\addtocounter{table}{-1}          

\begin{table}[tbh!]
\caption{
- continued$^{11}$.
\label{tab:crosscorr}
}
\begin{center}
\begin{tabular}{c c c c}
\hline
\hline
ID$_{MO}$ & ID$_{DM}$ & ID$_{BR}$ & Spectr. Ref. \\
\hline
     45464  & 29801$_2$   & 7096 & \\
     71299  & 30211$_3$   & - & \\
     21536  & 31067$_1$   & - & \\
         -  & 31087$_1$   & - & \\
         -  & 31090$_1$   & - & \\
     21554  & 31099$_1$   & 7041 & \\
     21555  & 31100$_1$   & 7045 & \\
     21558  & 31103$_1$   & - & \\
     21561  & 31106$_1$   & 7053 & \\
     21563  & 31108$_1$   & 7052 & \\
     21570  & 31117$_1$   & 7068 & \\
     21575  & 31122$_1$   & 7055 & \\
     21579  & 31127$_1$   & 7063 & \\
     21586  & 31134$_1$   & 7084 & \\
     21595  & 31143$_1$   & - & \\
     21596  & 31144$_1$   & - & \\
    187962  & 31292$_8$   & - & \\
         -  & 32759$_1$   & - & \\
     74452  &  4180$_4$   & - & \\
     50025  &  5668$_3$   & 7478 & \\
      4306  &  5891$_1$   & - & \\
         -  &  8420$_6$   & - & \\
    126376  &  8541$_6$   & - & \\
      6430  &  8864$_1$   & - & \\
    171895  &  9471$_8$   & - & \\
     23333  &   434$_2$   & 9836 & \\
\hline         
\end{tabular}
\end{center}
$^{11}$In Col. 1 we report the numeration of Montalto et al.~(2009), in 
Col. 2 the numeration of De Marchi et al. (2009), in Col. 3
the numeration of the WEBDA. In Col. 4 the reference for those
stars that were already spectroscopically studied by other authors. 
\end{table}

\newpage

\begin{figure*}
\center
\includegraphics[width=24cm,angle=90]{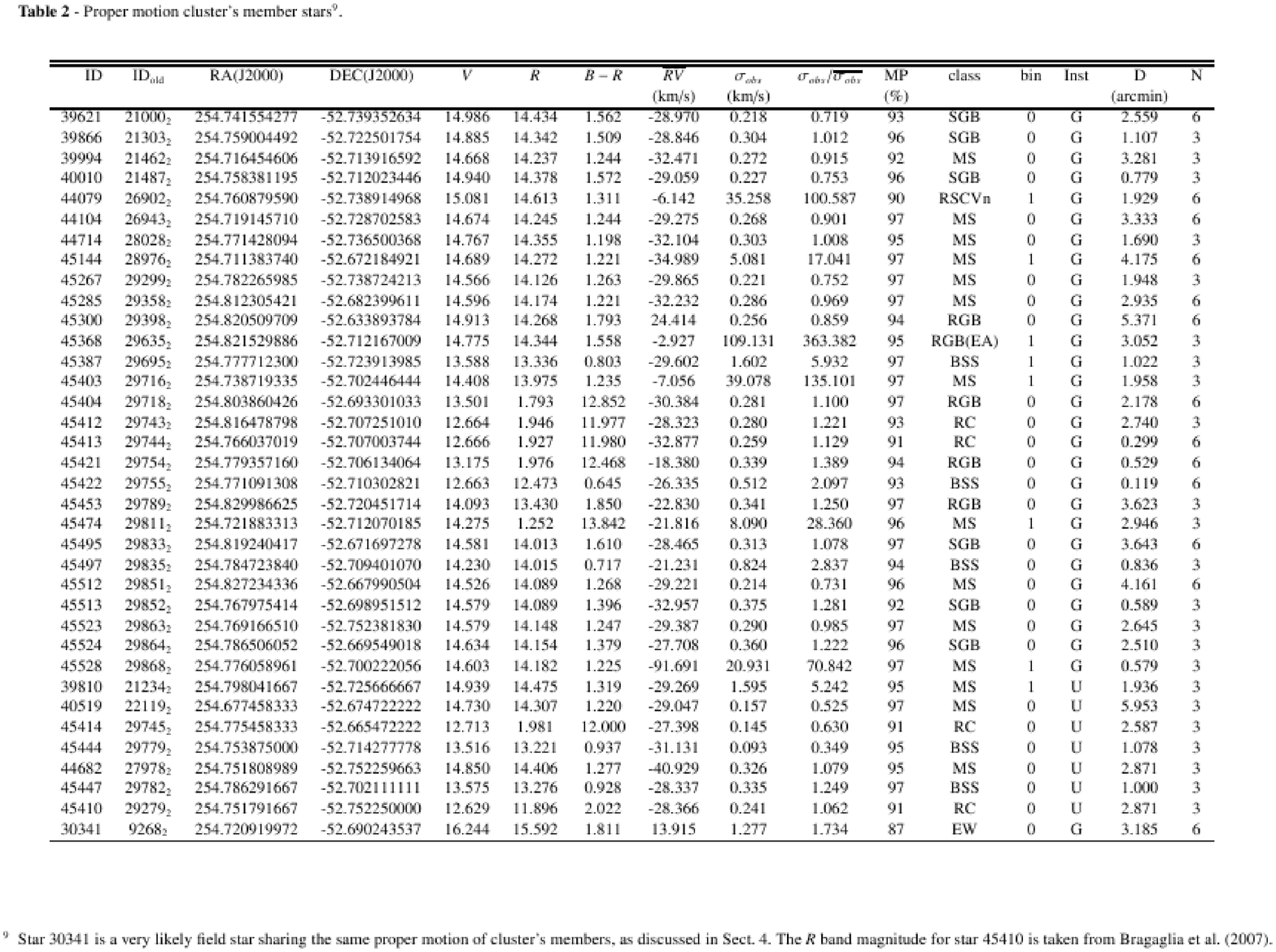}
\end{figure*}

\begin{figure*}
\center
\includegraphics[width=24cm,angle=90]{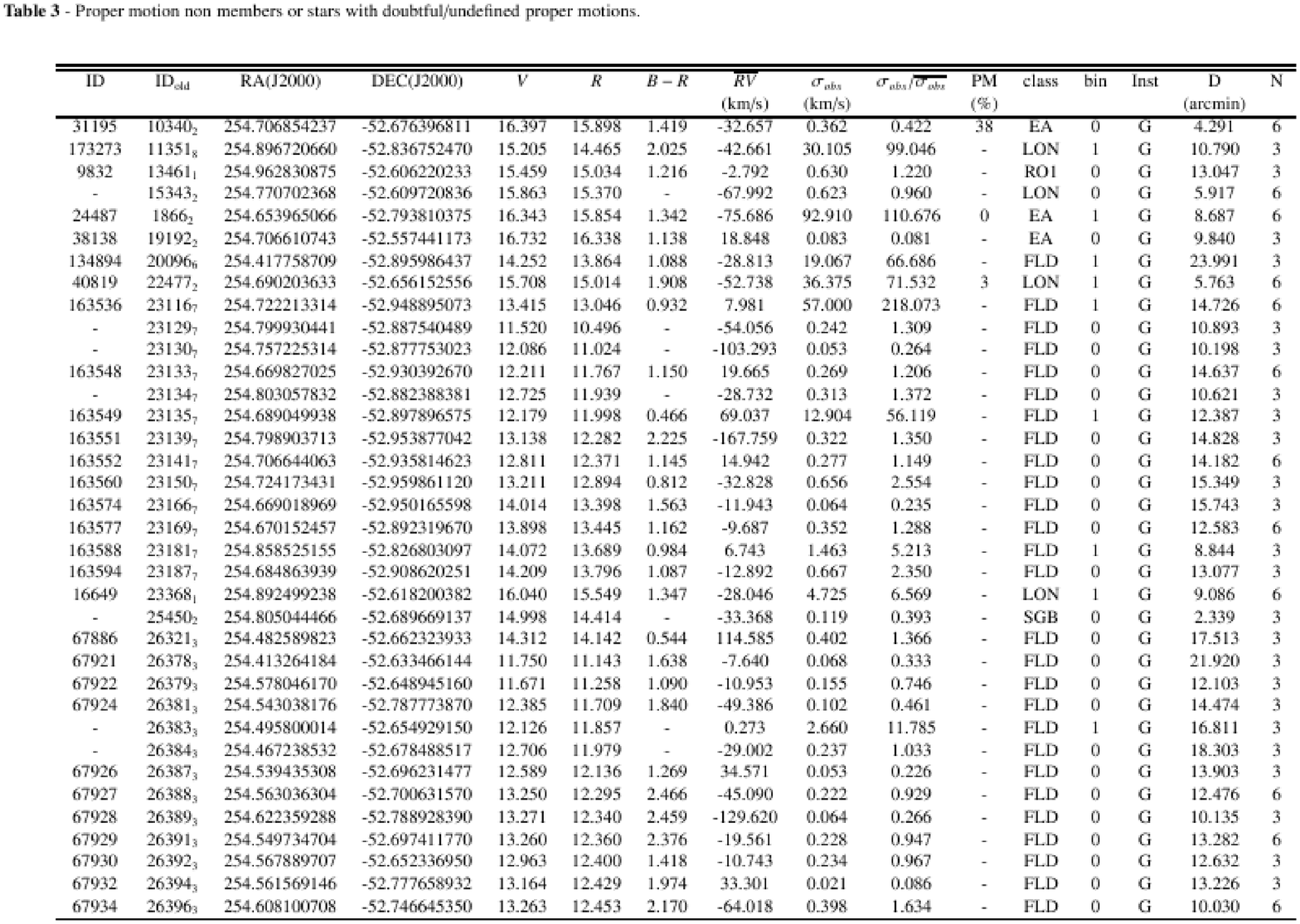}
\end{figure*}

\begin{figure*}
\center
\includegraphics[width=24cm,angle=90]{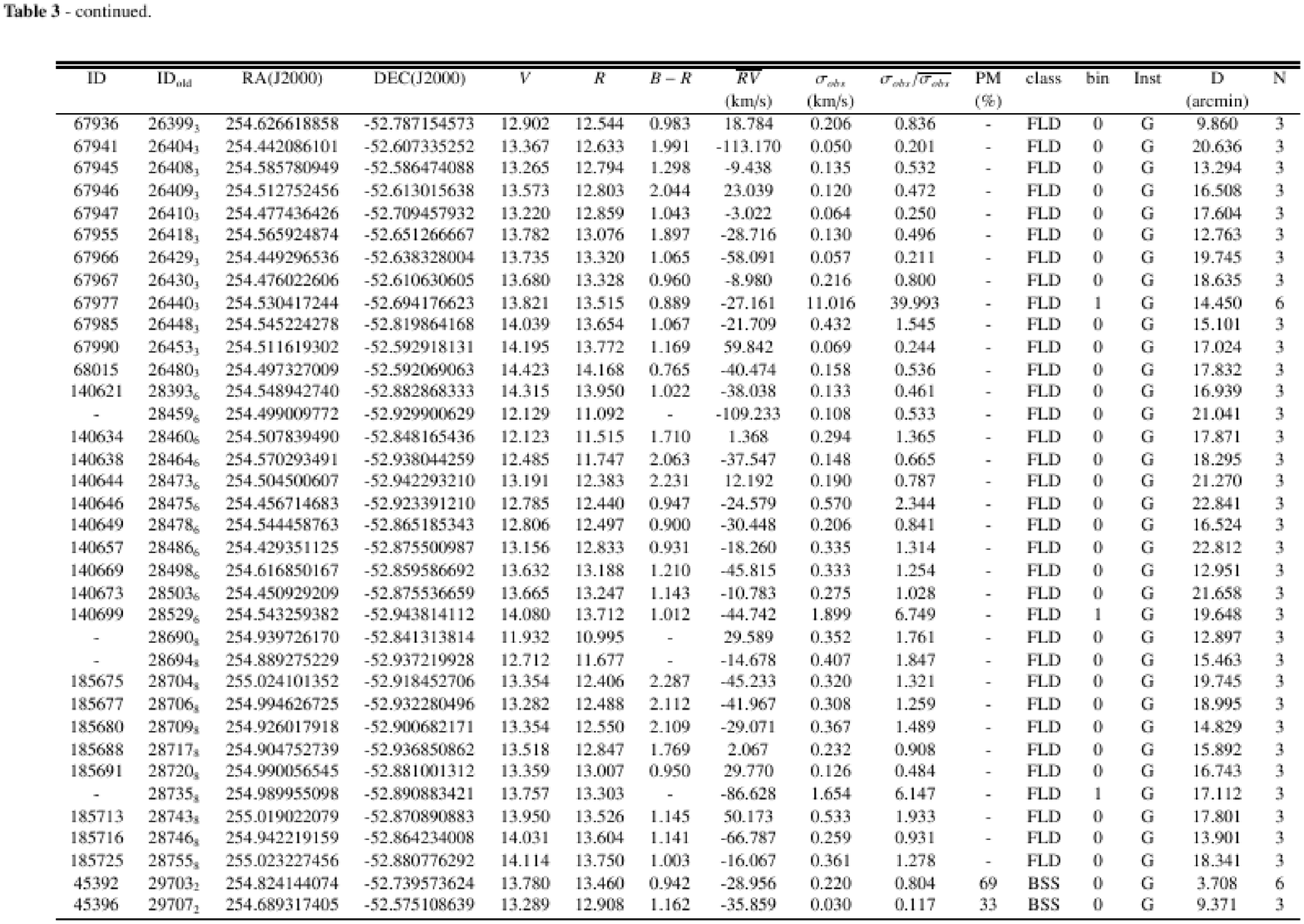}
\end{figure*}

\begin{figure*}
\center
\includegraphics[width=24cm,angle=90]{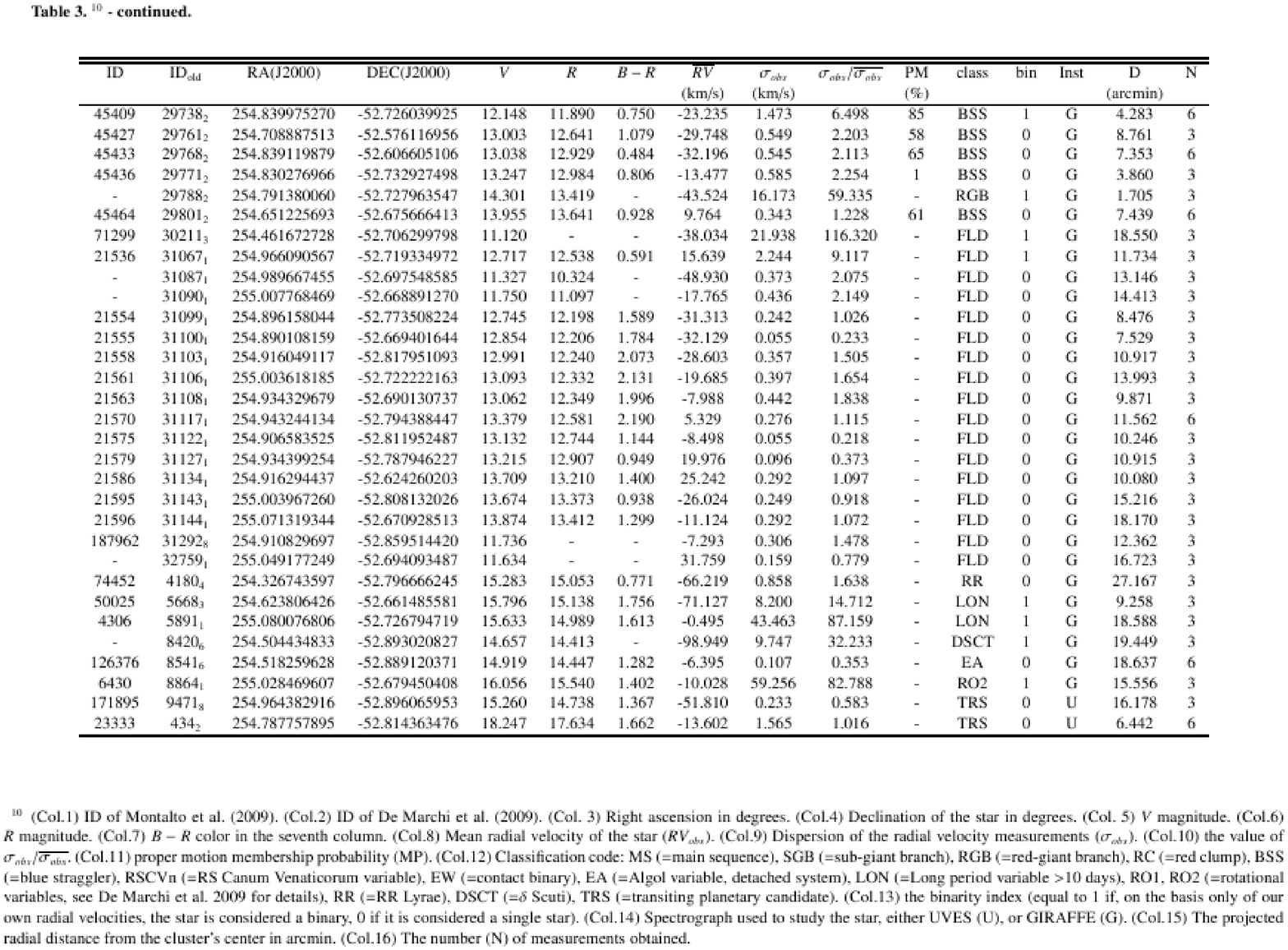}
\end{figure*}

\end{document}